\documentclass[aps,prd,eqsecnum,nofootinbib,twocolumn,showpacs,preprintnumbers,superscriptaddress,floatfix]{revtex4-1}

\usepackage[utf8]{inputenc}
\usepackage{amsmath,amssymb}
\usepackage{bm} 
\usepackage{CJK}
\usepackage{color}
\usepackage{dsfont} 
\usepackage{graphicx}
\usepackage{multirow}
\usepackage{subfiles}
\usepackage{tikz}

\usepackage[pdftex,colorlinks=true,
    linkcolor=blue,
    filecolor=blue,
    urlcolor=blue,
    citecolor=blue,
    plainpages=false,
    bookmarksopen=true]{hyperref} 

\newcommand{\order}{\ensuremath{\mathrm{O}}}

\newcommand{\onehalf}{\ensuremath{\frac{1}{2}}}

\newcommand{\Vlat}{V} 
\newcommand{\Alat}{A} 
\newcommand{\Jlat}{J} 
\newcommand{\Vcont}{\mathcal{V}} 
\newcommand{\Acont}{\mathcal{A}} 

\allowdisplaybreaks

\newcommand{\fermilab}{Fermi National Accelerator Laboratory, Batavia, IL 60510, USA}
\newcommand{\chicago}{Enrico Fermi Institute, University of Chicago, Chicago, IL 60637, USA}
\newcommand{\bnl}{Brookhaven National Laboratory, Upton, NY 11973, USA}
\newcommand{\IU}{Department of Physics, Indiana University, Bloomington, IN 47405, USA}
\newcommand{\lbnl}{present address: UC Berkeley and Lawrence Berkeley National Laboratory, Berkeley, CA 94720, USA}

\begin{document}

\begin{CJK*}{UTF8}{bsmi}
	
\title{Computing Nucleon Charges with Highly Improved Staggered Quarks}

\author{{Yin~\surname{Lin}} (林胤)}
\email{yin01@uchicago.edu}
\affiliation{\chicago}\affiliation{\fermilab}

\author{Aaron~S.~\surname{Meyer}}
\email{asmeyer.physics@gmail.com}
\altaffiliation{\lbnl}
\affiliation{\bnl}

\author{Steven~\surname{Gottlieb}}
\affiliation{\IU}

\author{Ciaran \surname{Hughes}}
\affiliation{\fermilab}

\author{{Andreas~S.~\surname{Kronfeld}}}
\email{ask@fnal.gov}
\affiliation{\fermilab}

\author{James~N.~\surname{Simone}}
\affiliation{\fermilab}

\author{Alexei~\surname{Strelchenko}}
\affiliation{\fermilab}

\collaboration{Fermilab Lattice Collaboration}
\noaffiliation

\date{\today}

\preprint{FERMILAB-PUB-20-551-T}

\begin{abstract}
This work continues our program of lattice-QCD baryon physics using staggered fermions for both the sea and valence quarks.
We present a proof-of-concept study that demonstrates, for the first time, how to calculate baryon matrix elements using staggered
quarks for the valence sector.
We show how to relate the representations of the continuum staggered flavor-taste group $\text{SU}(8)_{FT}$ to those of the discrete
lattice symmetry group.
The resulting calculations yield the normalization factors relating staggered baryon matrix elements to their physical counterparts.
We verify this methodology by calculating the isovector vector and axial-vector charges $g_V$ and $g_A$.
We use a single ensemble from the MILC Collaboration with 2+1+1 flavors of sea quark, lattice spacing $a\approx 0.12$~fm, and a pion
mass $M_\pi\approx305$~MeV.
On this ensemble, we find results consistent with expectations from current conservation and neutron beta decay.
Thus, this work demonstrates how highly-improved staggered quarks can be used for precision calculations of baryon properties, and,
in particular, the isovector nucleon charges.
\end{abstract}

\maketitle
\end{CJK*}

\section{Introduction}

Accurate first-principles calculations of nuclear cross sections are an important objective in the particle physics community.
In particular, heavy nuclei, such as $^{12}\text{C}$ and $^{40}\text{Ar}$, are used as targets in neutrino-scattering and
dark-matter detection experiments.
In calculations of cross sections, a necessary component is the modeling of nuclei as a collection of
nucleons, opening up an opportunity for lattice QCD~\cite{Kronfeld:2018qcd}.
At the quasielastic peak, for example, the electromagnetic and axial-vector form factors of the nucleon, which characterize the
electric charge and spin distribution within the nucleon, are key ingredients.
Such form factors can be obtained from the first-principles lattice-QCD framework.
However, these hadronic inputs remain one of the largest sources of systematic error as the experimental precision on these
cross-sections continues to improve~\cite{deAustri:2013saa,Alvarez-Ruso:2017oui,Ellis:2018dmb}.

The electromagnetic form factors have been extracted precisely from high statistics electron-nucleon scattering
experiments~\cite{Arrington:2006zm,Ye:2017gyb}.
At zero momentum transfer, the proton's electric form factor becomes the total electric charge $g_V=1$, and the slope at the origin
is related to the charge radius.
Recently, experiments that make use of the Lamb shift of muonic hydrogen report significantly smaller proton radii than those measured via
scattering~\cite{Lee:2015jqa}.
(For recent reviews of the proton radius puzzle, see Refs.~\cite{Hill:2017wzi,Hammer:2019uab}.) %
In addition, a recent reanalysis has demonstrated that the vector form factors at intermediate $Q^2$ also exhibit tensions outside
of their quoted uncertainties~\cite{Borah:2020gte}.
These disagreements could benefit from better knowledge of the Standard Model predictions, which necessitates using lattice QCD to
calculate the form factor.

In comparison, the nucleon axial-vector form factor is much less constrained from experimental data.
A recent re-analysis~\cite{Meyer:2016oeg} of the deuterium bubble-chamber data found greater uncertainties than previously assumed.
Again, lattice QCD can be illuminating here, computing the axial-vector form factor from first principles as an independent check on
the form factor extracted from experimental data.
At zero momentum transfer, the axial-vector form factor gives the so-called nucleon axial charge $g_A=1.2756(13)$, which has been
measured precisely in neutron beta decay~\cite{Zyla:2020zbs}.
Thus, the axial-charge can be used to validate lattice-QCD calculations before studying the momentum dependence of the form factor.
In addition, a percent-level first principles calculation of $g_A$ could shed light on the neutron lifetime
puzzle~\cite{Hill:2017wgb}.

Lattice-QCD calculations of baryonic observables are hindered by the well-known exponential growth of the noise relative to the
signal, which sets in at large times~\cite{Parisi:1983ae,Lepage:1989hd}.
At early times, where the signal-to-noise ratio is favorable, the lattice-QCD correlator data contain significant contributions
from several states in an infinite tower.
When using a fit to disentangle the higher-lying states from those of interest, some residual, unwanted contamination remains in
the parameters of interest.
It is imperative, therefore, to demonstrate control over both the noise and the excited-state contamination.

In this work, we use an ensemble generated by the MILC Collaboration~\cite{Bazavov:2012xda}, which incorporates a sea with
equal-mass up and down quarks, the strange quark, and the charm quark.
MILC uses the highly improved staggered-quark (HISQ) action~\cite{Follana:2006rc} for the sea quarks; here we use the HISQ action
for the valence quarks too.
Because staggered fermions have only one component per site and retain a remnant chiral symmetry, they are computationally
efficient.
Nevertheless, staggered fermions are complicated by the fermion doubling problem, leading to four species, known as tastes, for each
fermion field.
The four tastes become identical in the continuum limit, leading to an $\text{SU}(4n_f)$ flavor-taste symmetry for $n_f$ flavors.
Consequently, the spectrum of staggered lattice baryons is rich and intricate.
For nucleons, the spectrum has been classified~\cite{Golterman:1984dn,Bailey:2006zn,Lin:2019pia}, finding many
states that have the same properties as the physical nucleon.

In a recent paper, we used staggered baryons to calculate the nucleon mass~\cite{Lin:2019pia}.
Computing nucleon charges is the next step and a necessary one en route to the full momentum dependence of the form factors.
As discussed in Ref.~\cite{Lin:2019pia}, it can be advantageous to use unphysical nucleon-like states to carry out the calculation.
These states obtain the same properties as the physical nucleon in the continuum limit, where the full $\text{SU}(8)_{FT}$
flavor(isospin)-taste symmetry emerges.
For matrix elements such as charges and form factors, however, one must find the correct group-theoretic normalization factors
relating nucleon-like matrix elements to their physical counterparts.
This exercise is a straightforward if complicated application of the generalization of the Wigner-Eckart theorem to~SU(8).

To demonstrate this approach, we compute the nucleon vector and axial-vector charges on a single MILC HISQ ensemble with lattice
spacing $a\approx 0.12$~fm and pion mass $M_\pi\approx305$~MeV.
We employ local vector and axial-vector currents.
We also outline the steps needed to apply this method to matrix elements of other baryons, with an eye to future studies including
staggered baryons, such as $N\to\Delta$ transition form factors.

This paper is organized as follows.
In Sec.~\ref{sec:2pt3pt}, we discuss staggered-baryon correlators, starting with a brief review of the two-point correlator
methodology~\cite{Lin:2019pia}.
We then present an overview of our three-point correlators.
Here, we also present one of the key results of this paper: the correct normalization of the nucleon-like matrix elements.
In Sec.~\ref{sec:correlatorsmearing}, we describe strategies for removing excited-state contamination.
Section~\ref{sec:simdet} provides the details of our simulation, while Sec.~\ref{sec:fitting} describes Bayesian fits to the
correlator data.
Our computational results are presented in Sec.~\ref{sec:results}, including the robustness of our results under variations of our
fitting procedure, the renormalization of the bare lattice charge to the physical charges, and the final values for $g_V$ and $g_A$
on the single ensemble being used.
Finally, we compare our results to mixed-action results on the same ensemble and provide our conclusions in
Sec.~\ref{sec:conclusions}.
Appendices~\ref{app:matrix} and~\ref{app:WET} present the group theory relating the nucleon-like matrix elements to their physical
counterparts, including a numerical demonstration that these derivations are correct.

\section{Staggered Baryon Correlators}
\label{sec:2pt3pt}

For simplicity, we focus here on two flavors, up and down, with isospin symmetry.
With staggered fermions, instead of the usual $\text{SU}(2)_F$ isospin symmetry, an enlarged $\text{SU}(8)_{FT}$ flavor-taste
symmetry group emerges in the continuum limit.
It is important to note that the irreducible flavor-taste representations contain components with non-trivial taste and unexpected
isospin.
For example, Bailey has shown~\cite{Bailey:2006zn} that nucleon-like states exist with unphysical isospin yet masses equal in the
continuum limit to the physical tasteless nucleon.
In fact, all physics of such nucleon-like states can be related to that of the physical nucleon.
In particular, here we show how to relate nucleon-like matrix elements to their physical counterparts.
As such, we are allowed to choose any nucleon-like representation, for example, one that reduces the computational complexity.

We use the isospin-$\frac{3}{2}$ operators that transform in the $16$ irrep of the geometric timeslice group
(GTS)~\cite{Golterman:1984dn,Golterman:1985dz}, as presented in Ref.~\cite{Lin:2019pia}.
They are less complicated to analyze because only a single nucleon-like taste appears in the spectrum.
On the other hand, this irrep contains contributions from three $\Delta$-like tastes.

\subsection{Two-point correlators}
\label{sec:2pt}

Using the same notation as in Ref.~\cite{Lin:2019pia}, the two-point correlators read
\begin{align}
	C^{(r_1,r_2)}_\text{2pt} = \frac{1}{16}\sum_{s,\vec{D}}\sum_{\vec{x}}
        \left\langle B^{(r_2)}_{s\vec{D}}(\vec{x},t)\overline{B}^{(r_1)}_{s\vec{D}}(0)\right\rangle ,
    \label{eq:2ptraw}
\end{align}
using sink and source operators $B^{(r_2)}_{s\vec{D}}(\vec{x},t)$ and $\overline{B}^{(r_1)}_{s\vec{D}}(0)$ defined in 
Ref.~\cite{Lin:2019pia}.
To increase the statistical precision, we average over the eigenvalues $s=\pm$ of the staggered rotation in the $x$-$y$ plane, and
also the eight corners of the cube $\vec{D}$; together, $s$ and $\vec{D}$ label the components of the 16~irrep.
Here, $r_1, r_2 = 2,3,4,6$ represent four different operator constructions, or
``classes''~\cite{Golterman:1984dn,Bailey:2006zn,Lin:2019pia}, as well as other possible properties, such as smearing.

\subsection{Staggered-baryon matrix elements}
\label{sec:ME}

In this work, we are specifically interested in the isovector nucleon vector and axial-vector charges, namely $g_V$ and $g_A$,
respectively.
These are defined through the nucleon matrix elements
\begin{align}
	\langle N|\left(\bar{u}\Gamma_J u - \bar{d}\Gamma_J d\right)|N \rangle = g_J \, \bar{u}_N\Gamma_J u_N,
    \label{eq:gJ}
\end{align}
where $\Gamma_A =\gamma_z\gamma_5$ or $\Gamma_V = \gamma_4$, $u$ and $d$ are continuum-QCD up- and down-quark fields, and $u_N$ is
the nucleon spinor at zero momentum.

We calculate these nucleon matrix elements using (highly improved) staggered quarks.
To achieve this, we must extend the mass relations of Bailey~\cite{Bailey:2006zn} to matrix elements.
The baryon-like matrix elements and the physical matrix elements are related through symmetry transformations in the continuum.
In the appendices, we find the appropriate Clebsch-Gordan coefficients that relate the single-taste baryon matrix elements to the
physical tasteless QCD matrix elements by applying the generalized Wigner-Eckart theorem of~$\text{SU}(8)_{FT}$.

\begin{widetext}
The correctly normalized three-point correlators for our baryon-like operators are then
\begin{align}
    C_{V}^{(r_1,r_2)}(t,\tau) &= - \frac{1}{16}\sum_{\vec{D}}\sum_{\vec{x},\vec{y}} \mathcal{S}_V(\vec{D}) \left(
        \left\langle B^{(r_2)}_{-\vec{D}}(\vec{x},t) \Vlat(\vec{y},\tau) \overline{B}^{(r_1)}_{-\vec{D}}(0)\right\rangle + 
        \left\langle B^{(r_2)}_{+\vec{D}}(\vec{x},t) \Vlat(\vec{y},\tau) \overline{B}^{(r_1)}_{+\vec{D}}(0)\right\rangle \right),
	\label{eq:3ptraw_gV} \\
    C_{A}^{(r_1,r_2)}(t,\tau) &=   \frac{1}{16}\sum_{\vec{D}}\sum_{\vec{x},\vec{y}} \mathcal{S}_A(\vec{D}) \left(
        \left\langle B^{(r_2)}_{-\vec{D}}(\vec{x},t) \Alat(\vec{y},\tau) \overline{B}^{(r_1)}_{-\vec{D}}(0)\right\rangle - 3
        \left\langle B^{(r_2)}_{+\vec{D}}(\vec{x},t) \Alat(\vec{y},\tau) \overline{B}^{(r_1)}_{+\vec{D}}(0)\right\rangle \right),
    \label{eq:3ptraw_gA} 
\end{align}
where $t$ is the source-sink separation time and $\tau$ is the current insertion time.
The factor $-1$ in front of $C_V$ and the factor $-3$ in front of the second term of $C_A$ come from the group theory just
described.
Without these factors, these correlators would \emph{not} yield the desired nucleon charges.
For baryon operators and currents in other GTS irreps, different prefactors arise.
In Eqs.~(\ref{eq:3ptraw_gV}) and~(\ref{eq:3ptraw_gA}), we sum over unit-cube sites $\vec{D}$ with weights $\mathcal{S}_J(\vec{D})$
($J=V,A$) and have a separate term for each value of~$s=\pm1$.

Equations~(\ref{eq:3ptraw_gV}) and~(\ref{eq:3ptraw_gA}) introduce local currents
\begin{align}
    \Vlat(\vec{y},\tau) &= \mathcal{S}_{V}(\vec{y})\left(\bar{\chi}_u(\vec{y},\tau)\chi_u(\vec{y},\tau) -
        \bar{\chi}_d(\vec{y},\tau)\chi_d(\vec{y},\tau)\right),
    \quad
    \mathcal{S}_{V}(\vec{A}) = (-1)^{(A_x+A_y+A_z)/a},
    \label{eq:JV} \\
    \Alat(\vec{y},\tau) &= \mathcal{S}_{A}(\vec{y})\left(\bar{\chi}_u(\vec{y},\tau)\chi_u(\vec{y},\tau) - 
        \bar{\chi}_d(\vec{y},\tau)\chi_d(\vec{y},\tau)\right), 
    \quad
    \mathcal{S}_{A}(\vec{A}) = (-1)^{A_z/a},
    \label{eq:JA}
\end{align}
where $\chi_f$ is the field in the HISQ action of flavor~$f$.
The local vector and axial-vector currents, $\Vlat$ and $\Alat$, have spin-taste $\gamma_4\otimes\xi_4$ and
$\gamma_z\gamma_5\otimes\xi_z\xi_5$~\cite{Golterman:1985dz}, respectively.
The opposite-parity partners then arise from spin-taste $\gamma_5\otimes\xi_5$ and $\gamma_z\gamma_4\otimes\xi_z\xi_4$, respectively.
Because these local currents are not derived from Noether's theorem, they require a finite renormalization, that is, $Z_V\Vlat$ and
$Z_A\Alat$ have the same matrix elements as the continuum isovector currents in Eq.~(\ref{eq:gJ}).
\end{widetext}

In the limit $\tau\to\infty$ and $t-\tau\to\infty$, the ratio of the three-point to the two-point correlators approaches the desired
nucleon charge
\begin{align}
	\frac{C_{g_J}^{ (r_1,r_2)}(t,\tau)}{C^{(r_1,r_2)}_\text{2pt}(t)} \xrightarrow
        [t-\tau\rightarrow\infty]{\tau\rightarrow\infty} \widetilde{g}_J,
	\label{eq:3pt2pt}
\end{align}
where $\widetilde{g}_J$ is now the bare lattice charge, that is $g_J=Z_J\widetilde{g}_J$.
In practice, of course, we compute the correlators for several values of $t$ and $\tau$ and fit the $t$ and $\tau$ dependence to
extract the charges.

The finite renormalization factors $Z_J$ are determined first by noting that the remnant chiral symmetry requires
$Z_A=Z_V+\order(m_qa)^2$.
At zero momentum transfer, the vector current simply counts the number of up quarks minus the number of down quarks, for the proton
$2-1=1$.
One could, thus, define $Z_V$ by demanding $Z_V\widetilde{g}_V=g_V=1$.
Here, however, we prefer to define $Z_V$ via a similar relation obtained from a pseudoscalar-meson matrix
element~\cite{Gelzer:2019zwx}, then use the result to renormalize our nucleon matrix elements.
With this choice our result for $g_V$ is a genuine test of our methodology.

\section{Excited-State Contamination}
\label{sec:correlatorsmearing}

Excited-state contamination is one of the most difficult challenges when accurately estimating nucleon matrix elements from lattice
QCD.
The problem is even more complicated with staggered nucleons because of the presence of negative-parity and low-lying $\Delta$-like
states in the spectrum, which non-staggered formulations do not contain.
These are both significant sources of excited-state contamination in the present calculation.
We have, however, demonstrated control of excited-state contamination when extracting nucleon physics from two-point
staggered-baryon correlators~\cite{Lin:2019pia}.
Here we describe extensions of those techniques to the three-point correlators of the present work.
In particular, we show how to suppress contributions from the lowest-lying negative-parity states and the lowest three $\Delta$-like
tastes.

\subsection{Negative parity states}

Let $C_\text{2pt}(t)$ and $C_\text{3pt}(t,\tau)$ be any staggered-baryon correlators.
The source-sink separation is denoted $t$ and the current insertion time is denoted~$\tau$.
Any staggered operator that is local in time will create negative parity states, which in turn causes the characteristic
oscillations in time.
This is obvious from the correlators spectral decomposition
\begin{align}
    C_\text{2pt}(t)      &= z_+\bar{z}_+e^{-M_+t} + (-1)^{t/a} z_-\bar{z}_-e^{-M_-t} \nonumber \\
        & \hspace{0.75em} + \cdots , \\
    C_\text{3pt}(t,\tau) &= z_+A_{++}\bar{z}_+e^{-M_+t}  \nonumber\\
        & \hspace{0.75em} + (-1)^{t/a}        z_-A_{--}\bar{z}_-e^{-M_-t}\nonumber\\
        & \hspace{0.75em} + (-1)^{(t-\tau)/a} z_+A_{+-}\bar{z}_-e^{-M_+\tau}e^{-M_-{(t-\tau)}} \nonumber \\
        & \hspace{0.75em} + (-1)^{\tau/a}     z_-A_{-+}\bar{z}_+e^{-M_-\tau}e^{-M_+{(t-\tau)}} \nonumber \\
        & \hspace{0.75em} + \cdots,
    \label{eq:spectrasmearing}
\end{align}
where $M_\pm$ are the lowest-lying $\pm$ parity masses, $\bar{z}_\pm$ and $z_\pm$ are, respectively, the source and sink overlap
factors for states of parity $\pm$, and $A_{\pm\pm}$ and $A_{\pm\mp}$ are the transition matrix elements.
For simplicity, we have ignored backward propagating terms proportional to $e^{-M_\pm(T-t)}$, which are assumed to contribute
negligibly in the following.

Equation~(\ref{eq:spectrasmearing}) shows that the terms involving negative parity states change sign when either $t/a$ or $\tau/a$
change by one unit.
With this in mind, a time-averaging procedure can be applied to suppress the negative parity
contributions to the correlator.
A similar scheme was deployed in Ref.~\cite{Bailey:2008wp}.
The first ingredient is
\begin{align}
    C'_\text{2pt}(t) &= e^{-aM_\text{snk}}C_\text{2pt}(t) + C_\text{2pt}(t+a), \\
    C'_\text{3pt}(t,\tau) &= e^{-aM_\text{snk}}C_\text{3pt}(t,\tau) + C_\text{3pt}(t+a,\tau),
    \label{eq:smearsnk}
\end{align}
where we call $aM_\text{snk}$ the time-averaging parameter.
Substituting this expression into the spectral decomposition in Eq.~(\ref{eq:spectrasmearing}), one sees that the functional forms
of primed correlators are unchanged except that the sink overlap factors becomes
\begin{align}
    z_+ \to z_+ \left(e^{-aM_\text{snk}} + e^{-aM_+}\right), \\
    z_- \to z_- \left(e^{-aM_\text{snk}} - e^{-aM_-}\right).
\end{align}
If one chooses $aM_\text{snk} = aM_-$, then terms with the $M_-$ state at the sink will vanish, while the overlap factors for the
positive parity states become slightly larger.
In practice, the time-averaging parameter does not need to be exact to suppress the negative-parity states.

A similar time-averaging parameter, $aM_\text{src}$, can be introduced to reduce the negative parity contributions at the source via
\begin{align}
    C''_\text{2pt}(t)      &= e^{-aM_\text{src}}C_\text{2pt}(t)      + C_\text{2pt}(t+a),
    \label{eq:smearsrc} \\
    C''_\text{3pt}(t,\tau) &= e^{-aM_\text{src}}C_\text{3pt}(t,\tau) + C_\text{3pt}(t+a,\tau+a).
    \label{eq:smearsrc3}
\end{align}
Again, this step does not alter the functional forms of the two- and three-point correlators but replaces the source overlap
factors by
\begin{align}
    \bar{z}_+ &\to \bar{z}_+ \left(e^{-aM_\text{src}} + e^{-aM_+}\right) , \\
    \bar{z}_- &\to \bar{z}_- \left(e^{-aM_\text{src}} - e^{-aM_-}\right) .
\end{align}

If several negative parity states contribute significantly to the data, successive applications of this procedure, with suitable
parameters $[aM^{(1)}_\text{src},aM^{(2)}_\text{src}, \ldots]$ and $[aM^{(1)}_\text{snk},aM^{(2)}_\text{snk}, \ldots]$, can
appreciably suppress them.
On the other hand, because the relative error in the correlators becomes larger with time, too much time-averaging renders the data
statistically less precise.
Moreover, time-averaging reduces the available $\tau$ range in the modified correlators, thereby producing fewer data for the fit.
For each data set, some study is necessary to strike an optimal balance.

\subsection{\texorpdfstring{$\Delta$}{Delta}-like states}

Another source of excited-state contamination arises from the presence of the three $\Delta$-like states in the 16-irrep
correlators.
With four different classes of interpolators at both the source and the sink, we adopt the strategy from Ref.~\cite{Owen:2012ts} and
solve the generalized eigenvalue problem (GEVP)~\cite{Michael:1982gb,*Kronfeld:1989tb,*Luscher:1990ck}.
In Ref.~\cite{Lin:2019pia}, we applied the GEVP to our two-point correlators and successfully disentangled the nucleon-like state
from the $\Delta$-like states.
We extend that strategy to the three-point functions here.

Given a matrix two-point correlator, $\mathbf{C}_\text{2pt}(t)$, the left and right nucleon eigenvectors,
$\bm{u}(t_1,t_0)$ and $\bm{v}(t_1,t_0)$, are the solutions of
\begin{align}
    \mathbf{C}_\text{2pt}(t_1) \bm{u}(t_1,t_0) &= \lambda(t_1,t_0)\, \mathbf{C}_\text{2pt}(t_0)\bm{u}(t_1,t_0),
    \label{eq:eu} \\
    \bm{v}(t_1,t_0) \mathbf{C}_\text{2pt}(t_1) &= \lambda(t_1,t_0)\, \bm{v}(t_1,t_0) \mathbf{C}_\text{2pt}(t_0).
    \label{eq:ev}
\end{align}
Here, we focus on the eigenvectors for the nucleon-like state, the ones with the lowest eigenvalues, and put the others aside.
These eigenvectors optimize the projection onto the nucleon-like state in both the two- and three-point correlators via
\begin{align}
    C_\text{2pt}(t)      &= \bm{v}(t_1,t_0) \mathbf{C}_\text{2pt}(t)\bm{u}(t_1,t_0),
    \label{eq:projcorr2} \\
    C_\text{3pt}(t,\tau) &= \bm{v}(t_1,t_0) \mathbf{C}_\text{3pt}(t,\tau)\bm{u}(t_1,t_0).
    \label{eq:projcorr}
\end{align}
One has to decide which $t_1$ and $t_0$ to use in Eqs.~(\ref{eq:eu}) and~(\ref{eq:ev}).
The stability of our results under such variations will be discussed in Sec.~\ref{sec:simdet}.
Below we call the correlators in Eqs.~(\ref{eq:projcorr2}) and~(\ref{eq:projcorr}) the nucleon-optimized two- and three-point
correlators.

To summarize our strategy, we start with the correlators in Eqs.~(\ref{eq:2ptraw}), (\ref{eq:3ptraw_gA}), and~(\ref{eq:3ptraw_gV}),
and apply two iterations of time-averaging at both the source and sink, and then project the time-averaged correlation matrix as in
Eq.~(\ref{eq:projcorr}).
The time-averaging suppresses the negative parity states contributions, and the projection suppresses the $\Delta$-like baryons
contributions.

\section{Simulation Details}
\label{sec:simdet}

To demonstrate the feasibility of nucleon matrix elements with staggered quarks, we use a single gauge ensemble, which was generated
by the MILC collaboration~\cite{Bazavov:2012xda}.
MILC implemented the one-loop, tadpole-improved L\"uscher-Weisz gauge action~\cite{Hart:2008sq}, as well as the HISQ
action~\cite{Follana:2006rc} for the sea, which contains equal-mass up and down quarks, the strange quark, and the charm quark.
In this work, we also employ the HISQ action for the valence quarks, with the same mass as the up-down sea quarks.

The ensemble has dimension $L^3\times T=24^3\times64$, a lattice spacing $a=0.1222(3)$~fm (determined from the $F_{p4s}$
mass-independent scheme~\cite{Bazavov:2017lyh}), a pion mass $M_\pi\approx305$~MeV, and a light-to-strange-quark mass ratio of
$1/5$.
Other parameters of this ensemble are listed in Ref.~\cite{Bazavov:2017lyh}.
Note that the CalLat~\cite{Berkowitz:2017gql,*Chang:2018uxx} and the PNDME~\cite{Gupta:2018qil} collaborations have both used this
same ensemble to calculate $g_A$, albeit with either the M\"obius domain wall or Wilson-clover valence fermion actions,
respectively.

We generate the two- and three-point correlators according to Eqs.~(\ref{eq:2ptraw}), (\ref{eq:3ptraw_gV}),
and~(\ref{eq:3ptraw_gA}).
We measure each correlator on $872$ configurations, and further increase the statistics by randomly placing the corner-wall sources
on eight maximally separated timeslices to give a total of $6976$ measurements per correlator.
(The $\tau/a=7$ correlators have only four time sources per configuration.)

We block all measurements in a single gauge configuration and every four consecutive gauge trajectories to avoid autocorrelations.
The covariance matrix between different correlator components are estimated with the non-linear shrinkage method \cite{ledoit2012}
to avoid ill-conditioning from finite sample sizes.

As described in Ref.~\cite{Lin:2019pia}, we use corner-wall sources to optimize the signal-to-noise ratio and point sinks.
In the present work, we remove the Coulomb-gauge fixed links, as we have empirically observed that leaving out the links has little
effect on correlators but with the added advantage of a simpler code.
Here we also incorporate the Wuppertal smearing~\cite{Gusken:1989qx,Bali:2016lva} at the sink by applying
\begin{align}
    \chi^{(n)} &= \left(1+\frac{3\sigma^2}{32a^2N}\Delta\right)\chi^{(n-1)},
    \label{eq:wup} \\
    \Delta \chi(\vec{x}) &= -6\chi(\vec{x}) +
        \sum_{i=1}^{3}\left[\chi(x_i+2a) + \chi(x_i - 2a)\right]
    \label{eq:wup2} 
\end{align}
in order to reduce excited state contamination.
In Eq.~(\ref{eq:wup}), $n$ is the $n^{\text{th}}$ iteration of $N$ total iterations; all shifts are stride~2 to preserve the
staggered symmetries.
We include the appropriate gauge transporters to make the smearing gauge covariant~\cite{Lin:2019pia}, but for succinctness they are
omitted from Eq.~(\ref{eq:wup2}).
We generate data with two different root-mean-squared (rms) smearing radii, $\sigma$, which are about $0.2$ and $0.6$~fm.
We label the two smearings as Gr2.0N30 and Gr6.0N70.

For the three-point correlators, we invert the propagators from the current insertion to obtain all operator classes at the sink.
Five current insertion times, $\tau/a = [3,4,5,6,7]$, are generated for both the vector and axial-vector current insertions.

To eliminate the unwanted negative parity states, we then pass all the correlators through two iterations of time-averaging using
Eqs.~(\ref{eq:smearsnk}) and (\ref{eq:smearsrc}) with
\begin{align}
	[aM^{(1)}_\text{src}, aM^{(2)}_\text{src}]=[aM^{(1)}_\text{snk}, aM^{(2)}_\text{snk}] = [0.9, 1.1].
    \label{eq:srcsnksmearparam}
\end{align}
These two numbers are based on an observation in Ref.~\cite{Lin:2019pia} that the lowest-lying negative parity state seems to have
energy around the S-wave $N\pi$ state, which in this ensemble is about $0.9$ in lattice units.
We then set the second averaging parameters about $aM_\pi\sim 0.2$ higher than the first ones, which again is consistent with our
findings in Ref.~\cite{Lin:2019pia}.
As the goal is to suppress the negative-parity states, the accuracy of these parameters is not crucial.
Note that each iteration of the source time-averaging in Eq.~(\ref{eq:smearsrc3}) reduces the current insertion timeslices by one,
so the time-averaged three-point correlators have only $\tau/a=[3,4,5]$.
The smearing in Eq.~(\ref{eq:smearsnk}), on the other hand, does not reduce the range for~$\tau/a$.

\begin{figure}
    \centering
    \includegraphics[width=\columnwidth]{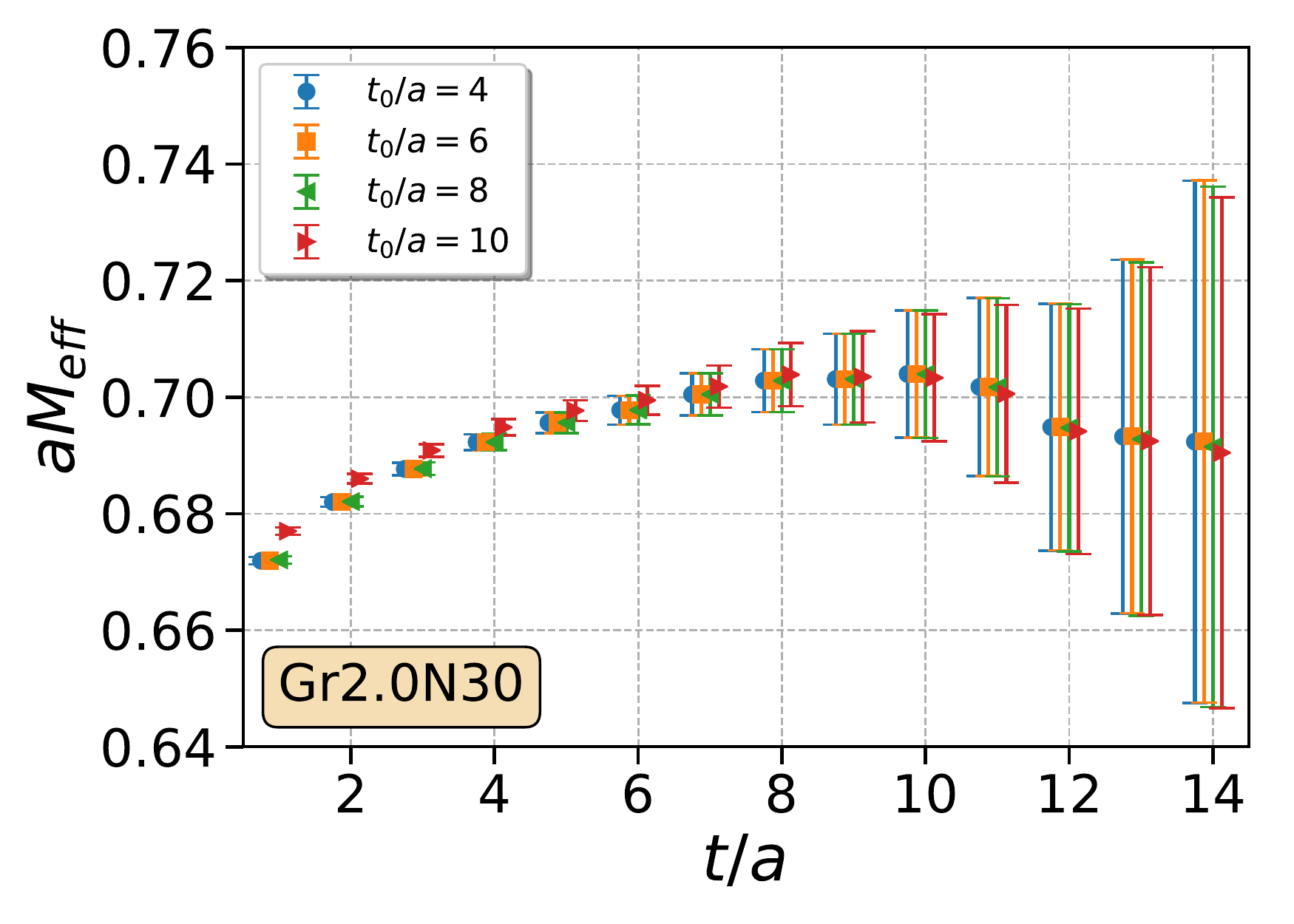}
    \includegraphics[width=\columnwidth]{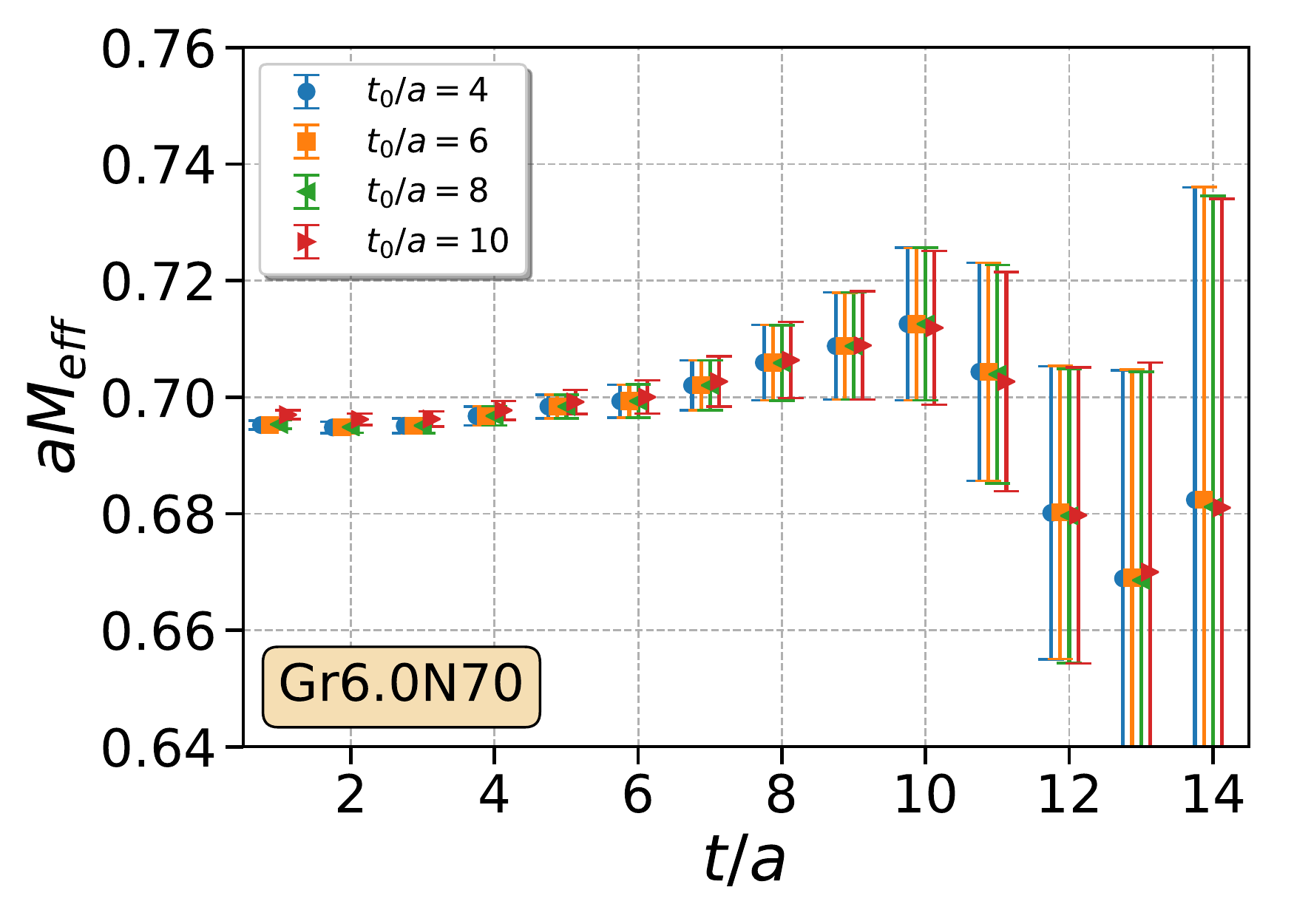}
    \caption{(Color online) 
        Effective masses of the nucleon-optimized correlators as a function of the source-sink separation time~$t$.
        The top plot has Wuppertal sink smearing radius $\sigma_\text{rms} = 0.2$~fm (Gr2.0N30), and the bottom
        $\sigma_\text{rms}=0.6$~fm (Gr6.0N70).
        The time-averaging parameters are given in Eq.~(\ref{eq:srcsnksmearparam}).
        Different colored points represent different choices of $t_0$ when solving the GEVP equation in Eq.~(\ref{eq:ev}),
        as shown in the legends, and are offset slightly for clarity.}
    \label{fig:2pt_vary_t0}
\end{figure}

After time-averaging, we solve for the left and right eigenvectors using Eq.~(\ref{eq:ev}) in order to optimize our correlators as
in Eq.~(\ref{eq:projcorr}).
To ensure the robustness of our fitting methodology, we test the stability of our results under variations of the choice of $t_0$.
To do so, we compute the effective mass of the optimized two-point correlators, which we define as
\begin{align}
    aM_\text{eff}(t) \equiv \frac{1}{2}\ln\left(\frac{C_\text{2pt}(t)}{C_\text{2pt}(t+2a)}\right).
\label{eq:em}
\end{align}
The $t_0$ stability plots are shown in Fig.~\ref{fig:2pt_vary_t0}.
All choices produce similar results, and so we choose $t_0=6a$ for the subsequent analyses.
Similarly, we vary $(t_1-t_0)/a$ from $2$ to $6$ and find, again, that the differences are negligible, so we fix $t_1-t_0=2a$.

\begin{figure*}
    \centering
    \includegraphics[width=\columnwidth]{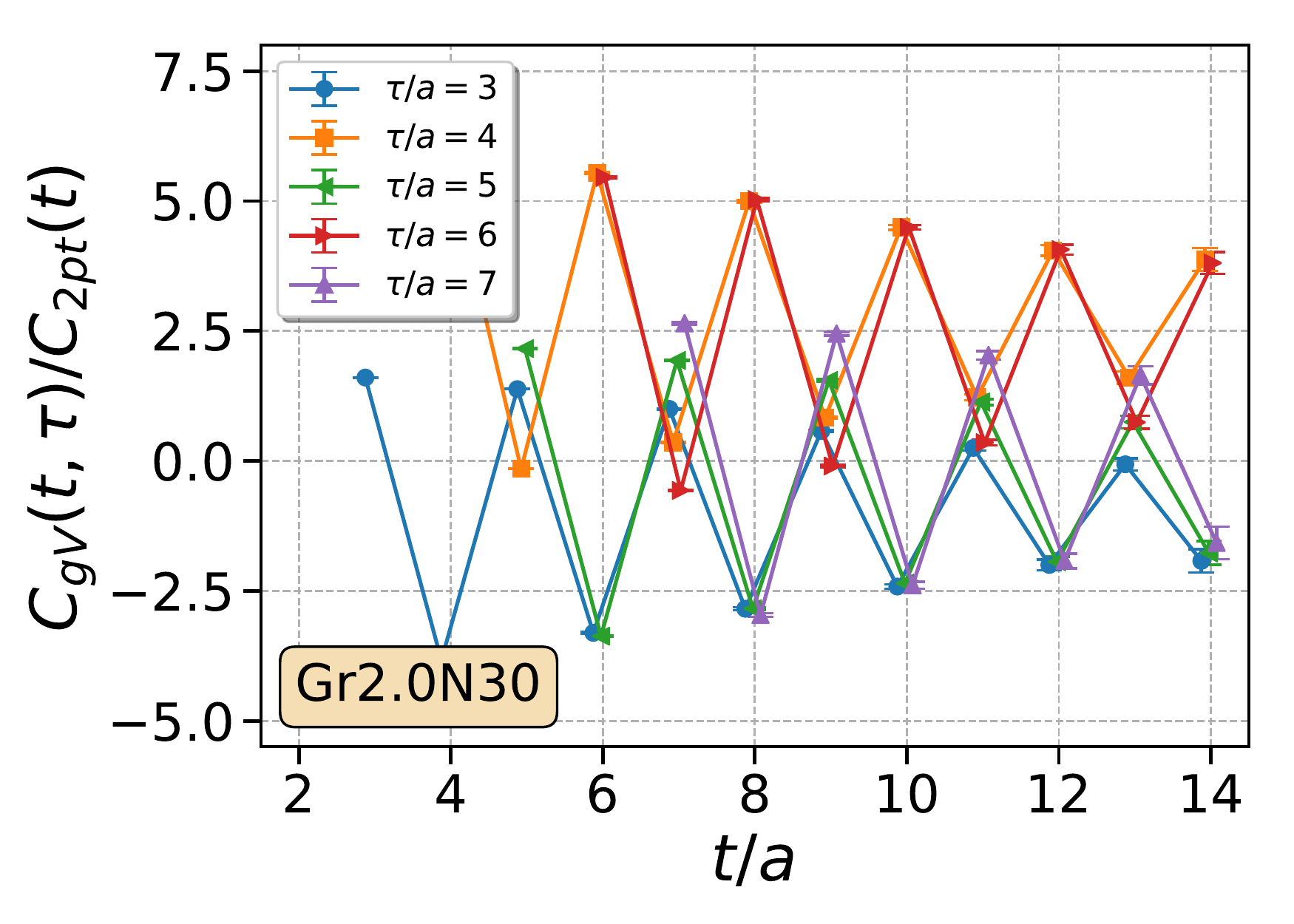} \hfill
    \includegraphics[width=\columnwidth]{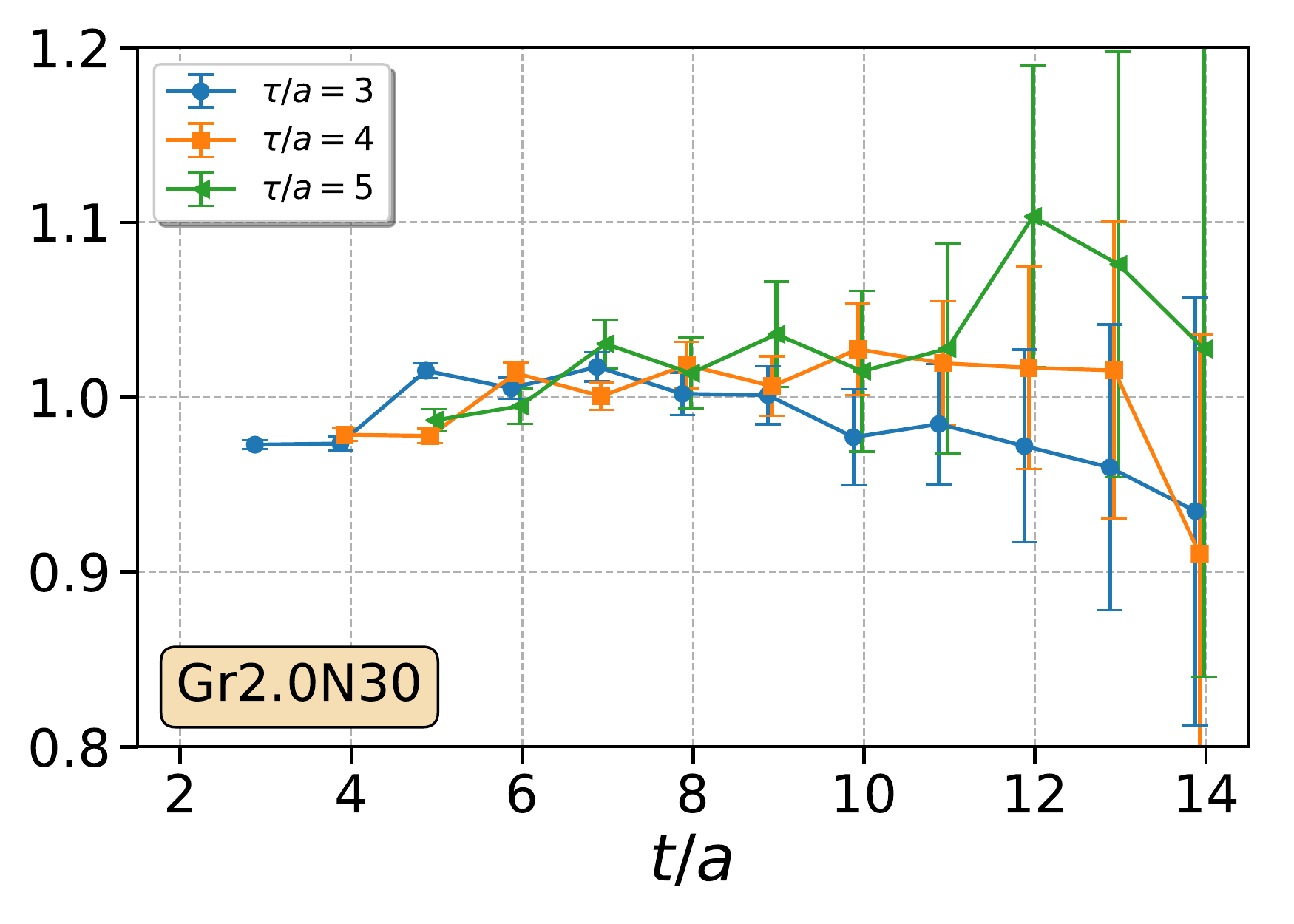}   \\
    \includegraphics[width=\columnwidth]{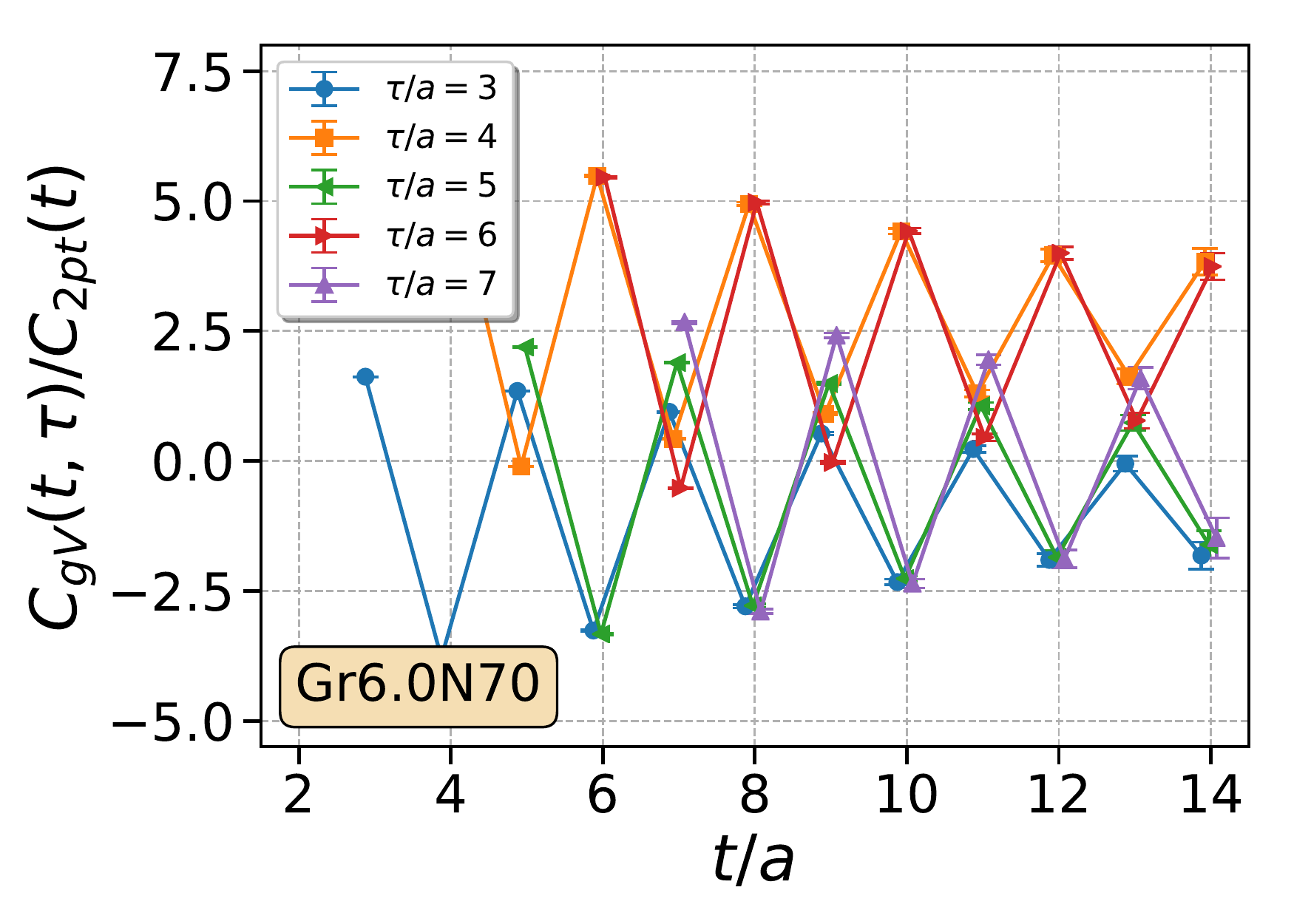} \hfill
    \includegraphics[width=\columnwidth]{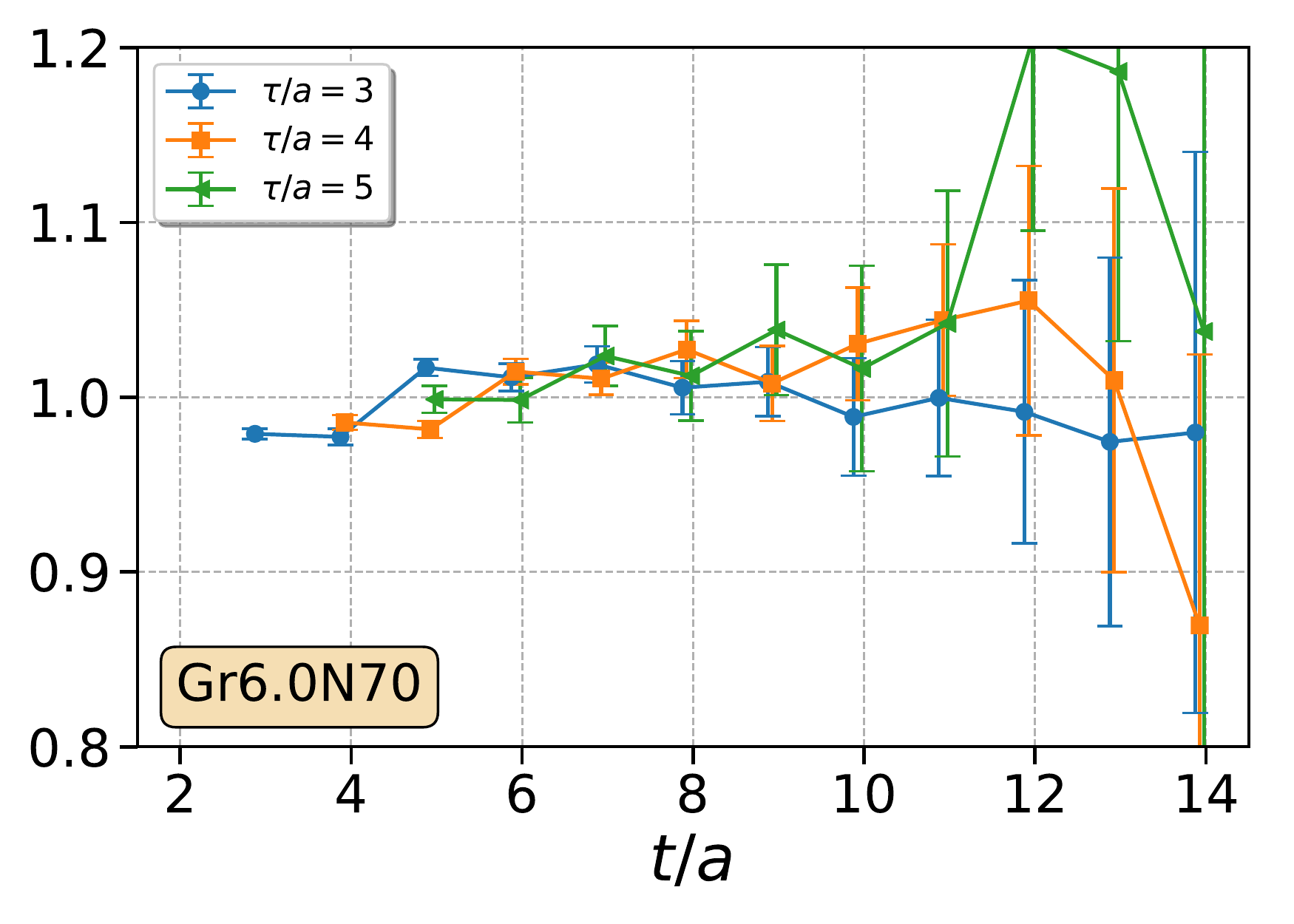}
    \caption{(Color online) %
    The $g_V$ three-point to two-point nucleon-optimized correlator ratio as a function of the source-sink
    separation time $t$, and current insertion time~$\tau$.
    Correlators are labeled by the rms Wuppertal smearing radii, $\sigma_\text{rms}=0.2$~fm (Gr2.0N30) and
    $\sigma_\text{rms}=0.6$~fm (Gr6.0N70).
    In the limits $\tau,t-\tau\rightarrow \infty$, this ratio converges to the bare $g_V$ nucleon charge.
    The correlators in the left column are not time-averaged with the oscillation suppressing procedure described in
    Sec.~\ref{sec:correlatorsmearing}.
    The right column shows data that are time-averaged with the parameters given in Eq.~(\ref{eq:srcsnksmearparam}), yielding
    much smoother curves reminiscent of non-staggered fermion correlators.}
    \label{fig:gvvarysmear}
\end{figure*}

\begin{figure*}
    \centering
    \includegraphics[width=\columnwidth]{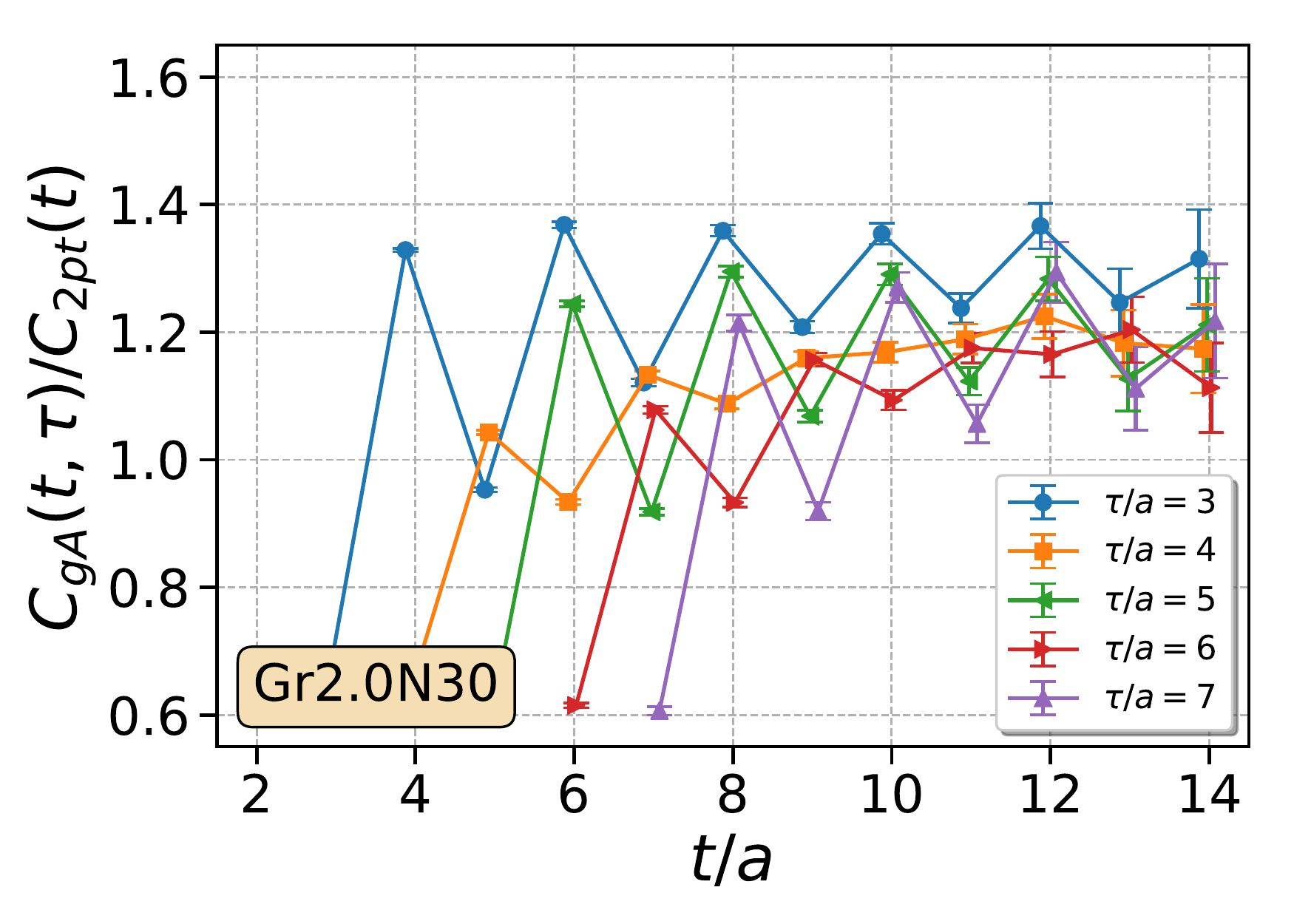} \hfill
    \includegraphics[width=\columnwidth]{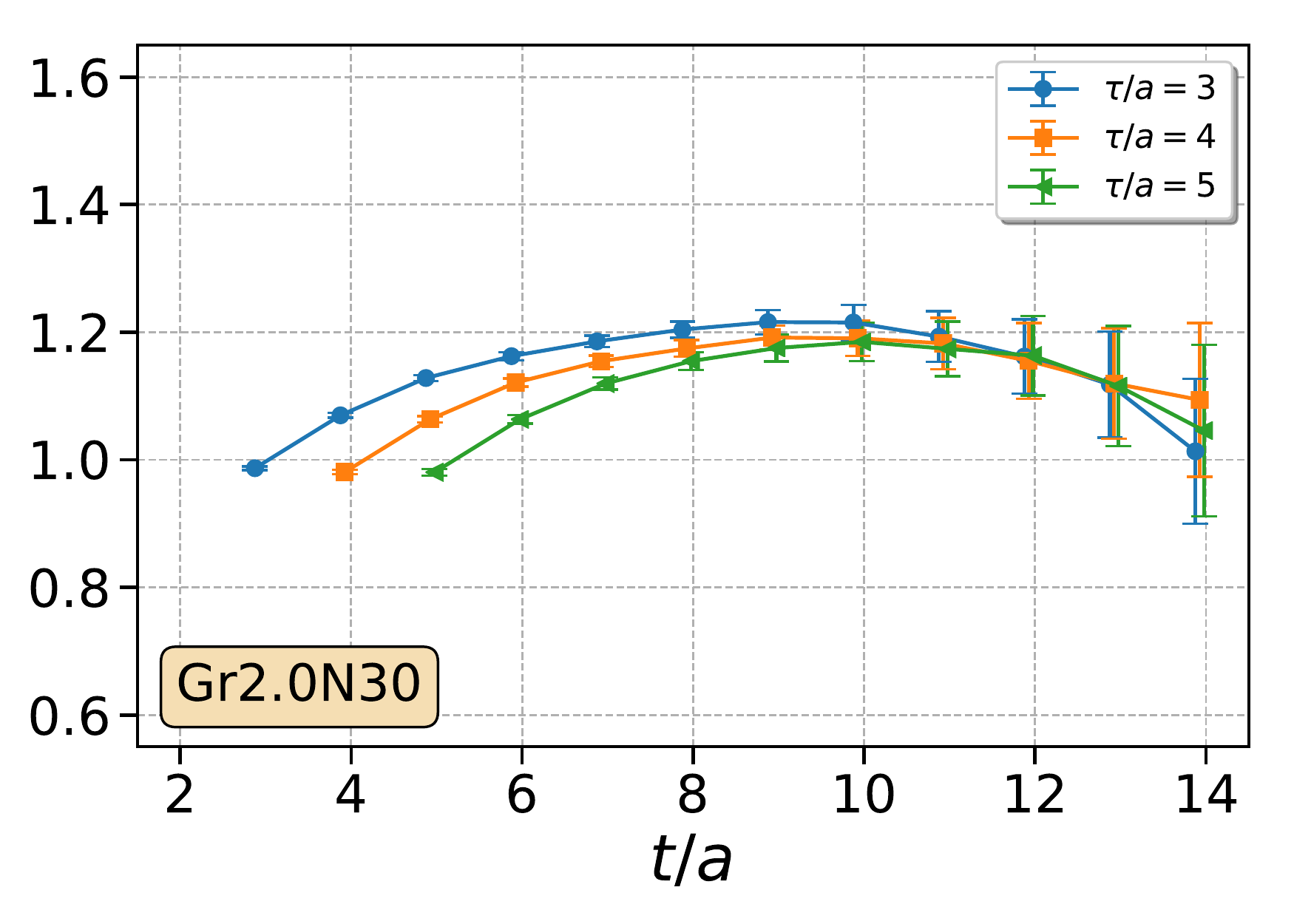}   \\
    \includegraphics[width=\columnwidth]{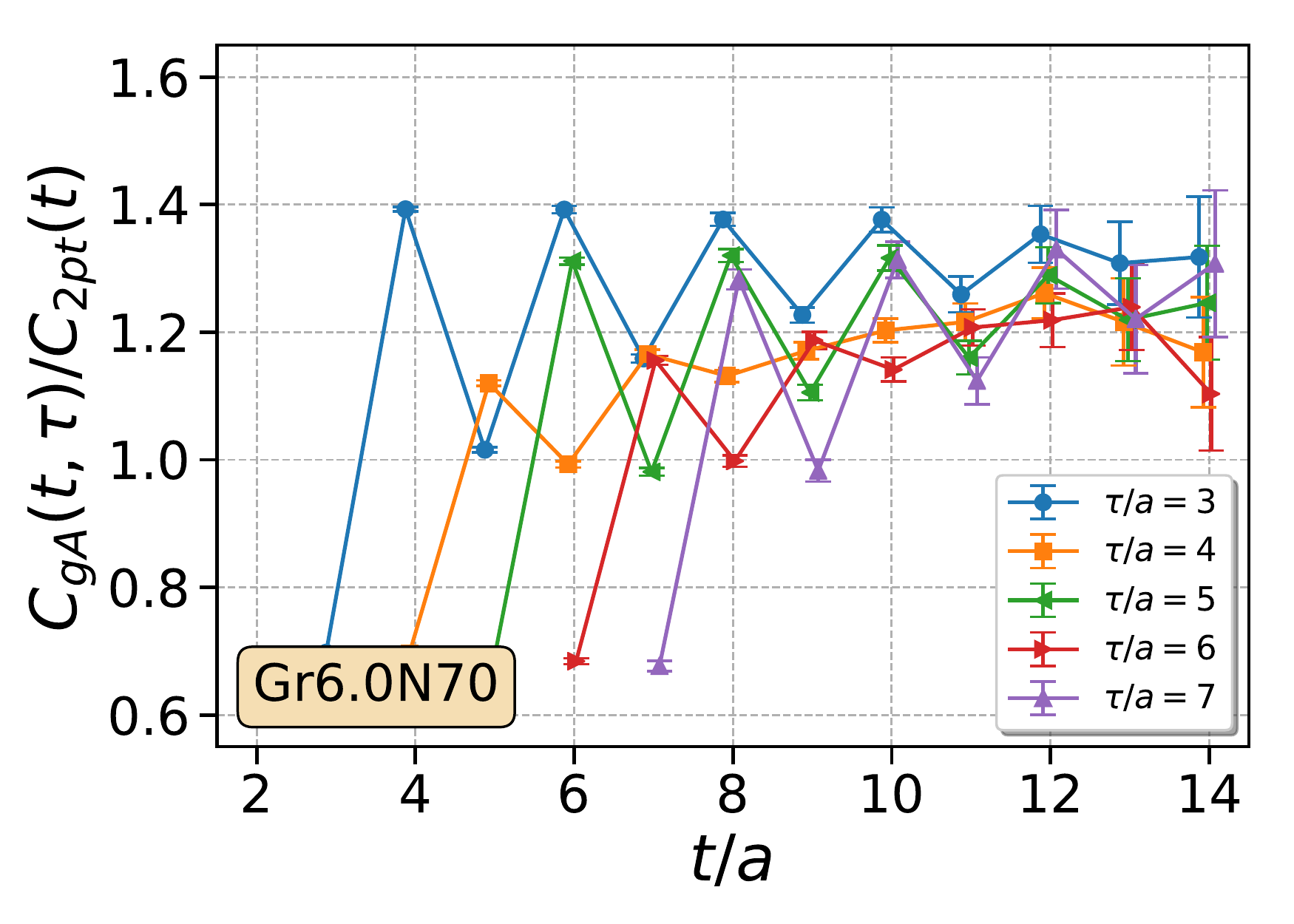} \hfill
    \includegraphics[width=\columnwidth]{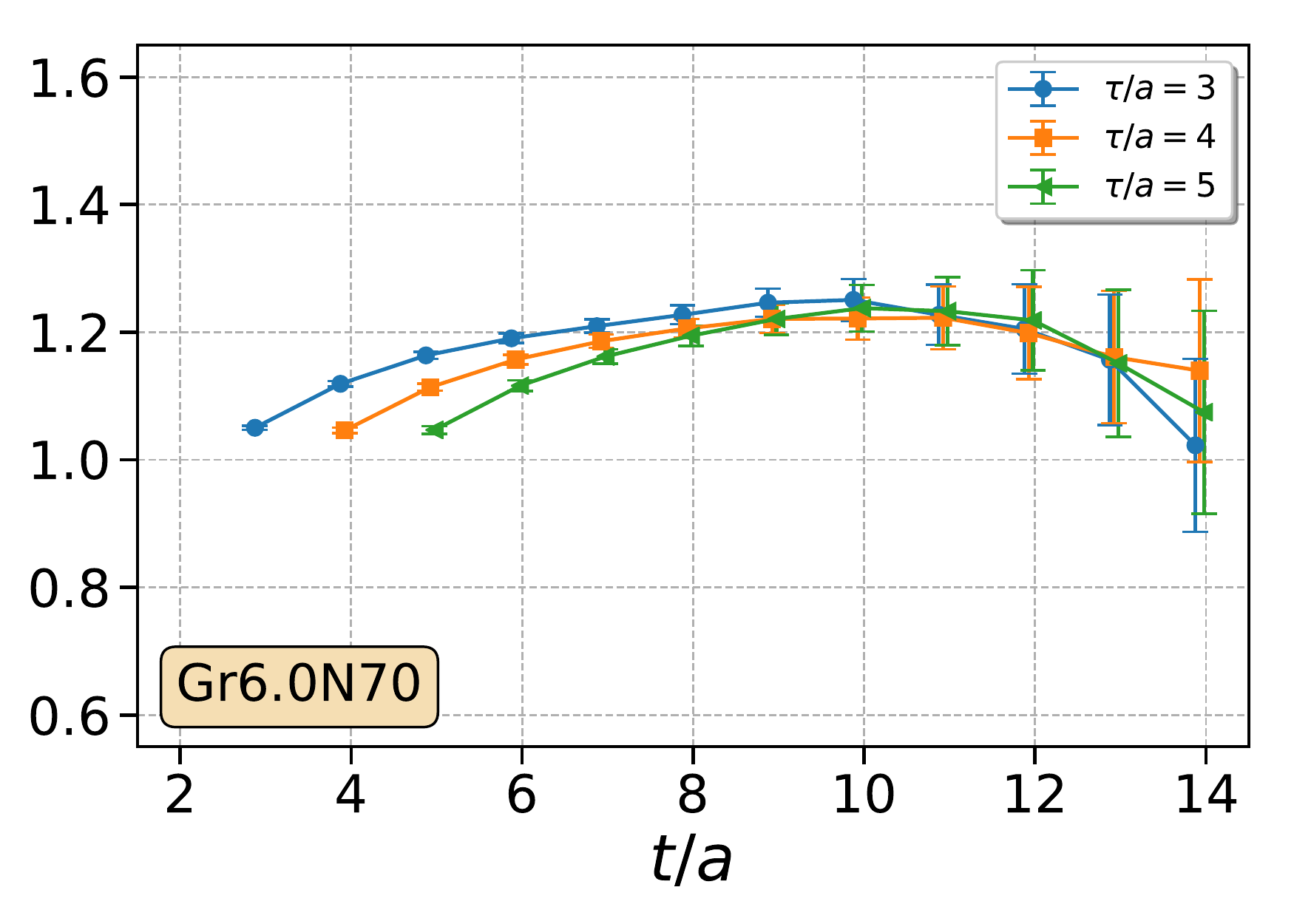}
    \caption{(Color online) %
        Identical to Fig.~\ref{fig:gvvarysmear} but with $g_A$ instead of $g_V$.
        See the caption of Fig.~\ref{fig:gvvarysmear} for further details.}
    \label{fig:gavarysmear}
\end{figure*}

Since we have normalized the nucleon-like three-point correlators correctly, and each of the correlator transformations that we
perform preserve the functional form of the spectral decomposition, the optimized three-to-two-point correlator ratios converge to
the desired nucleon charges in the large-time limits.
In Figs.~\ref{fig:gvvarysmear} and~\ref{fig:gavarysmear}, we plot the ratio of the nucleon-optimized three-to-two-point correlators,
with and without time-averaging at the source and sink.
The left column shows the optimized correlators without time-averaging, and the right column shows them time-averaged with the
parameters given in Eq.~(\ref{eq:srcsnksmearparam}).
The two smearing are shown in the top (Gr2.0N30) and bottom (Gr6.0N70) rows.
Significant oscillations are clearly present in the unaveraged correlators, particularly for the vector current in
Fig.~\ref{fig:gvvarysmear}.
This is expected, because the parity partner of the vector current is the pseudoscalar current $P$, and the
$\langle N\pi|P|N\rangle$ matrix element gives a large contribution to the vector-current data and, thus, causes large oscillations.
For the $g_A$ data, on the other hand, the parity partner of the axial current is the tensor current $T_{z4}$, and when the nucleon
is at rest $\langle N|T_{z4}|N\rangle=0$.
Consequently, the first non-zero contribution in the axial-vector parity partner channel will likely be from
$\langle N\pi|T_{z4}|N\rangle$, leaving small oscillations.

\section{Correlator Fitting}
\label{sec:fitting}

We apply the Bayesian fitting methodology implemented in \texttt{corrfitter}~\cite{peter_lepage_2020_3890467} to extract the nucleon
mass and matrix elements.
We observe in Fig.~\ref{fig:gvvarysmear} that the vector correlators have noticeable oscillatory contributions, whereas the
axial-vector correlators shown in Fig.~\ref{fig:gavarysmear} do not.
Further, the vector correlators seem relatively insensitive to our choice of Wuppertal smearing.
We perform separate fits to the vector and axial-vector correlators, but include their correlations through bootstrapping.

It should be stressed that, after applying the excited-state suppression techniques from Sec.~\ref{sec:correlatorsmearing}, the
interpretation of the higher exponentials in the correlators is ambiguous.
For the positive-parity channel, the first ``excited state'' could be a mixture of any leftover $\Delta$-like states, the P-wave
$N\pi$ states, or other finite volume energy levels higher up in the spectrum that are related to resonances.
For the negative-parity channel, we found in Ref.~\cite{Lin:2019pia} that the ground state is likely to contain S-wave $N\pi$
states.
The time-averaging procedure to cancel out the negative-parity states makes identification of these states even more ambiguous.
Regardless of the origin, we can treat the excited states as nuisance parameters and fit them away with an exponential fit function.
In this case, each excited exponential mass parameter describes a conglomeration of several eigenstates of the Hamiltonian.
Still, we will refer to each exponential in the fit function as a state without necessarily identifying it with any single
eigenstate.
As discussed in Ref.~\cite{Lin:2019pia}, the stability of the extracted fit parameters as a function $t_\text{min}$ indicates lack
of excited-state contamination, as long as they are modeled accurately.
This $t_\text{min}$ stability plot is shown in Fig.~\ref{fig:stab}, and is discussed in Sec.~\ref{sec:fitstab}.

\subsection{Functional forms for fitting}
\label{sec:fit}

\begin{table}
    \centering
    \caption{Summary of the prior choices for the fit Ans\"atze given in Eqs.~(\ref{eq:2ptans}), (\ref{eq:gAans}),
        and~(\ref{eq:gVans}).
        All prior distributions are Gaussian, except for $M_{+1}-M_{+0}$, which is log-normal.}
    \label{tab:prior} 
    \setlength{\tabcolsep}{6pt}
    \begin{tabular}{lr@{ $\pm$ }l}
        \hline\hline
        Quantity & Prior value & width\\
        \hline
        $M_{+0} = \text{nucleon mass}$ & 1100 & 200~MeV\\
        $M_{+1}-M_{+0}$                &  300 & 200~MeV\\
        $M_{-0}$                       & 1600 & 300~MeV\\
        $A_{+0,+0} = \widetilde{g}_A$  & 1.2 & 0.3 \\
        $V_{+0,+0} = \widetilde{g}_V$  & 1.0 & 0.3 \\
        $A_{i,j}; i\neq+0, j\neq +0$   & 0.0 & 5.0 \\
        $V_{i,j}; i\neq+0, j\neq +0$   & 0.0 & 5.0 \\
        \hline\hline
    \end{tabular}
\end{table}

For the $g_A$ analysis, we perform simultaneous fits to the optimized two- and three-point correlators, and include both the
Gr2.0N30 and Gr6.0N70 sink smearings.
Observation of the strong suppression of excited states in Fig.~\ref{fig:gvvarysmear} leads us to use a fit ansatz that contains two
positive-parity states and one negative-parity state:
\begin{align}
    C^{\sigma,\text{fit}}_\text{2pt}&(t) =
        z^{\sigma}_{+0}\bar{z}_{+0}e^{-M_{+0}t} + z^{\sigma}_{+1}\bar{z}_{+1}e^{-M_{+1}t} \nonumber \\
        &+(-1)^{t/a}z^\sigma_{-0}\bar{z}_{-0}e^{-M_{-0}t},
    \label{eq:2ptans} \\
    C^{\sigma, \text{fit}}_A&(t,\tau) = 
        \sum_{i,j=0}^{1}z^\sigma_{+i}A_{+i, +j}\bar{z}_{+j}e^{-M_{+i}\tau}e^{-M_{+j}(t-\tau)}\nonumber\\
        &+z^\sigma_{-0}A_{-0, -0}\bar{z}_{-0i}(-1)^{t/a}e^{-M_{-0}t}
    \label{eq:gAans} \\
        &+\sum_{i=0}^{1}z^\sigma_{-0}A_{-0, +i}\bar{z}_{+i}(-1)^{(t-\tau)/a}e^{-M_{+i}\tau}e^{-M_{-0}(t-\tau)}\nonumber\\
        &+\sum_{i=0}^{1}z^\sigma_{+i}A_{+i, -0}\bar{z}_{-0}(-1)^{\tau/a}e^{-M_{-0}\tau}e^{-M_{+i}(t-\tau)}. \nonumber
\end{align} 
Here, $M_{+0}= M_N$ is the nucleon mass, $M_{+1}$ is the mass of the first residual positive-parity excited state, $\bar{z}_{+i}$
and $z^\sigma_{+i}$ are their source and sink overlap factors (with sink smearing $\sigma=0.2$, $0.6$~fm), $M_{-0}$ is the mass of
the residual negative-parity state, and $\bar{z}_{-0}$ and $z^\sigma_{-0}$ the source and sink overlap factors.
The terms $A_{\pm i, \pm j}$ are the unrenormalized axial-vector matrix elements, with $A_{+0,+0}=\widetilde{g}_A$ the desired bare
axial charge.
Note that the two-point correlator terms involving finite temporal $T$ extent are not included here since we average our data
symmetrically around the $T/2$ point as described in Ref.~\cite{Lin:2019pia}.

In the Bayesian fit, we choose Gaussian priors for the ground-state masses, overlap factors, and matrix elements.
We choose log-normal priors for the mass differences between adjacent states to enforce the ordering of states.
It has been observed empirically that the nucleon mass has an approximate linear dependence on the pion mass (see, for example,
Ref.~\cite{Walker-Loud:2014iea}), so we choose a prior of $1100\pm 200$~MeV for the nucleon mass on our ensemble with
$M_\pi=305$~MeV.
We put a wide prior of $300\pm200$~MeV centered at the $\Delta$-like mass for the mass splitting $M_{+1} - M_{+0}$ to
accommodate for potential mixing of many physical states.
For the same reason, we also impose a wide mass prior of $1600\pm300$~MeV for the negative-parity mass $M_{-0}$ centered at the
S-wave $N\pi$~state.
All prior choices are summarized in Table~\ref{tab:prior}.

\emph{A priori}, we have no knowledge of the sign or magnitude of the overlap factors.
Consequently, all overlap factors are effectively unconstrained.
Very wide priors of $0\pm5$ are chosen for all matrix elements, apart from $\widetilde{g}_A=A_{+0,+0}$, for which we choose a wide
prior of $1.2\pm0.3$ centered near the PDG~\cite{Zyla:2020zbs} value of~$g_A$.
As discussed below in Sec.~\ref{sec:results}, we know from other work with pseudoscalar mesons that $Z_A$ is close enough to unity
not to influence the choice of prior.

For the $g_V$ analysis, we use the same two-point functional form as Eq.~(\ref{eq:2ptans}).
However, for the three-point correlators we use
\begin{align}
    C^{\sigma, \text{fit}}_{g_V}&(t,\tau) = \sum_{i=0}^{1} z^\sigma_{+i}V_{+i,+i}\bar{z}_{+i}e^{-M_{+i}t}
    \nonumber \\
        &+z^\sigma_{-0}V_{-0, -0}\bar{z}_{-0i}(-1)^{t/a}e^{-M_{-0}t}
    \label{eq:gVans} \\
        &+\sum_{i=0}^{1}z^\sigma_{-0}V_{-0, +i}\bar{z}_{+i}(-1)^{(t-\tau)/a}e^{-M_{+i}\tau}e^{-M_{-0}(t-\tau)} \nonumber \\
        &+\sum_{i=0}^{1}z^\sigma_{+i}V_{+i, -0}\bar{z}_{-0}(-1)^{\tau/a}e^{-M_{-0}\tau}e^{-M_{+i}(t-\tau)}, \nonumber
\end{align} 
where the notation is identical to that of Eq.~(\ref{eq:gAans}).
The $V_{i,j}$ are the unrenormalized vector matrix elements.
The $V_{+i,+j}$ with $i\neq j$ are omitted on the first line of Eq.~(\ref{eq:gVans}), because they are forbidden by
vector charge conservation, up to small discretization effects.
The priors are also identical to the $g_A$ fits except for the bare vector charge, $\widetilde{g}_V$.
Given that the renormalization constant is close to unity, we choose the $\widetilde{g}_V$ prior to be $1.0\pm0.3$.

\subsection{Fit Stability}
\label{sec:fitstab}

The most important part of the nucleon matrix element fitting procedure is separating the nucleon observables of interest from the
excited-state contributions.
To demonstrate the lack of excited-state contamination, we examine the stability of the observables as choices in the fit are
varied.
Specifically, we vary $t_\text{min}$, $\Delta\tau_\text{min}$, and $t_\text{max}$ where $t_\text{min}$ is the minimum source-sink
separation time that we include in our two-point correlator fits, $\Delta\tau_\text{min}$ is the minimum source-sink separation time
\emph{after} the current insertion time, $\tau$, that we include in our three-point correlator fits, and $t_\text{max}$ is the
maximum source-sink separation time.
The nominal parameters for the nominal fits are given in Table.~\ref{tab:param}.

\begin{table}
    \centering
    \caption{Summary of the nominal fit range parameters. $t$ is the source-sink separation time, and $\tau$ is the current
        insertion time.
        $t_\text{min}$ and $t_\text{max}$ are the minimum and maximum source-sink separation included in the nominal fits;
        $\Delta\tau_\text{min}$ is the minimum time \emph{after} the current insertion time that we include in the three-point
        fits.}
    \label{tab:param} 
    \setlength{\tabcolsep}{6pt}
    \begin{tabular}{lcc}
       \hline\hline
       Correlator & Fit Parameter & Nominal value\\
       \hline
       \multirow{2}{*}{Two-point}&$t_\text{min}/a$&~5\\
       &$t_\text{max}/a$& 13\\
       \hline
      \multirow{2}{*}{Three-point} & $\Delta\tau_\text{min}/a$&~3\\
      &$t_\text{max}/a$& 13\\
       \hline\hline
    \end{tabular}
\end{table}

\begin{figure}
    \centering
    \includegraphics[width=0.98\columnwidth]{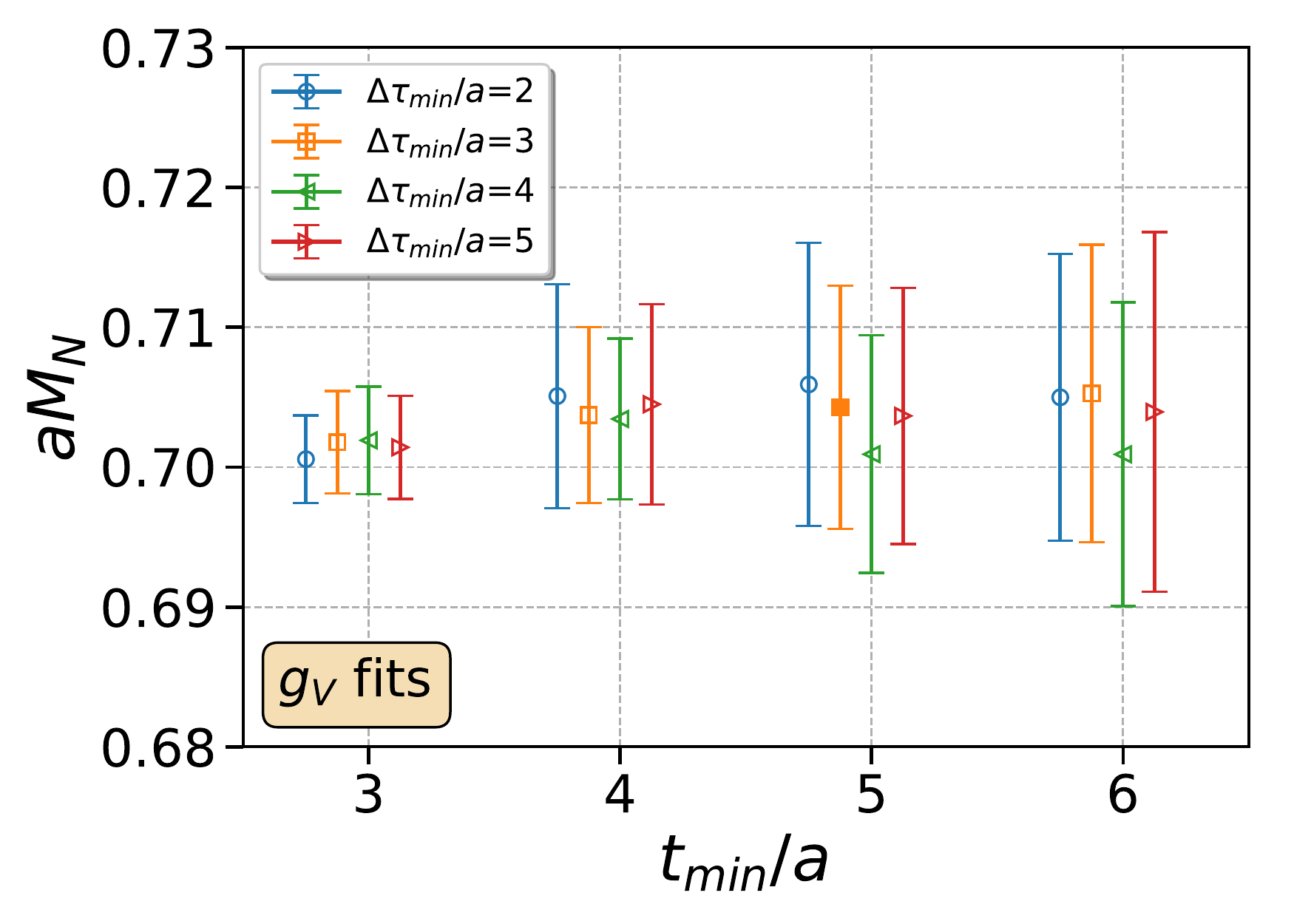}
    \includegraphics[width=0.98\columnwidth]{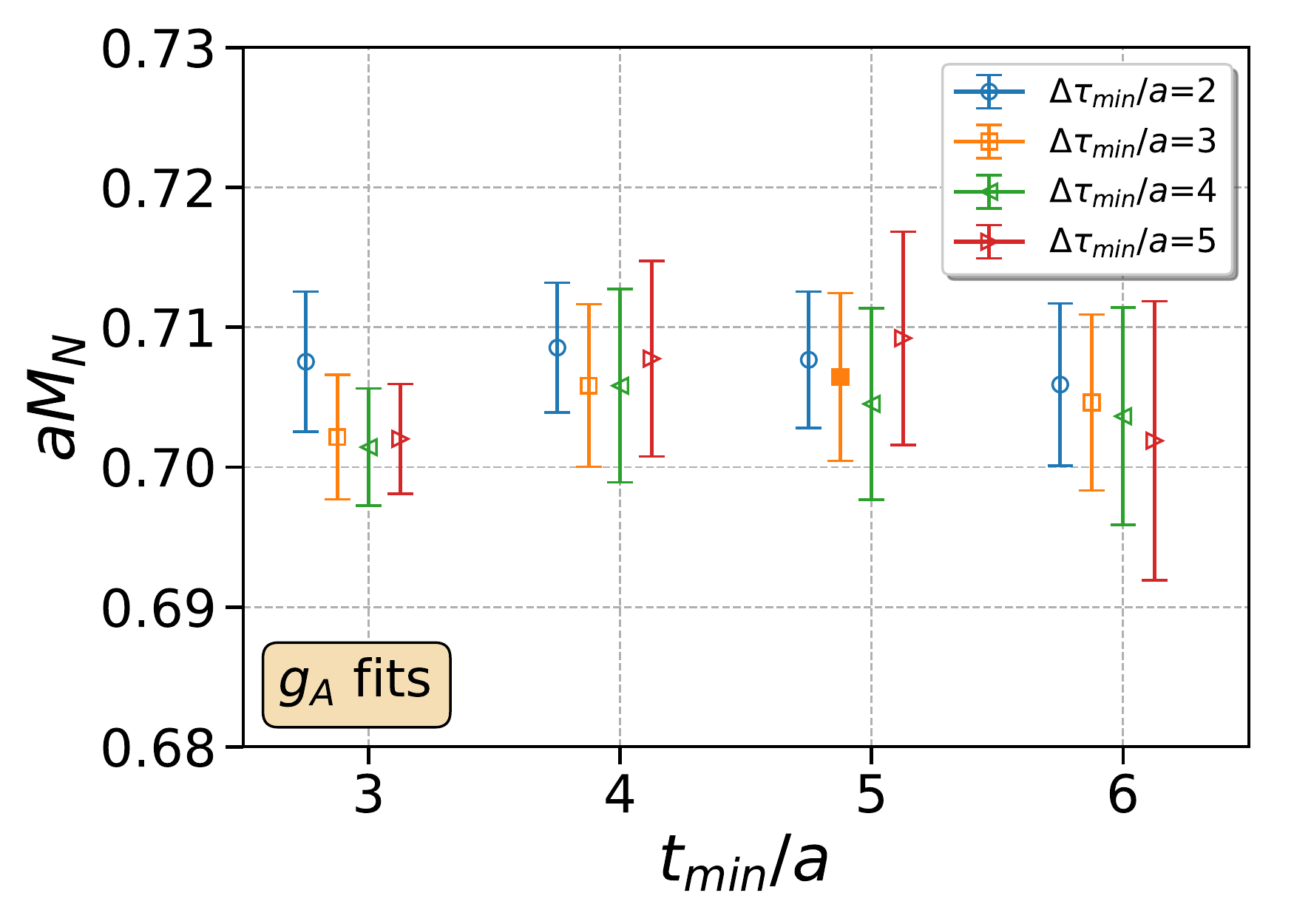}
    \caption{(Color online) %
        The stability plot for the extracted nucleon mass, $aM_N$, as a function of $t_\text{min}$ and
        $\Delta\tau_\text{min}$, obtained from either the $g_V$ (top) or $g_A$ (bottom) fits.
        The definitions of $t_\text{min}$ and $\Delta\tau_\text{min}$ are described in the text.
        The maximum source-sink separation time is fixed at $t_\text{max}=13a$ for all correlators.
        The solid squares are the nominal fit results, and all uncertainties are estimated with $1000$ bootstrap samples.}
    \label{fig:stab_M}
\end{figure}

\begin{figure}
    \centering
    \includegraphics[width=0.98\columnwidth]{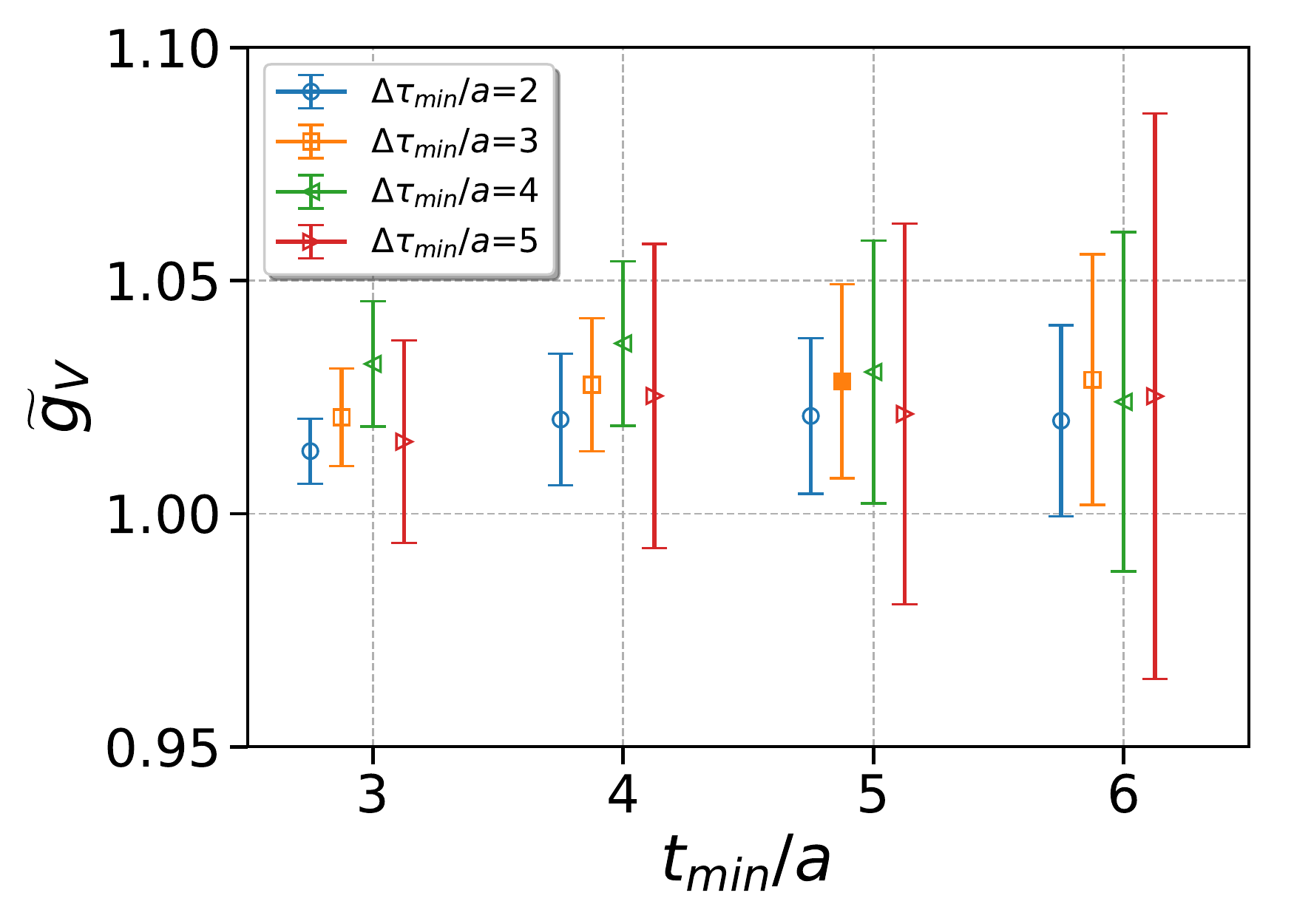}
    \includegraphics[width=0.98\columnwidth]{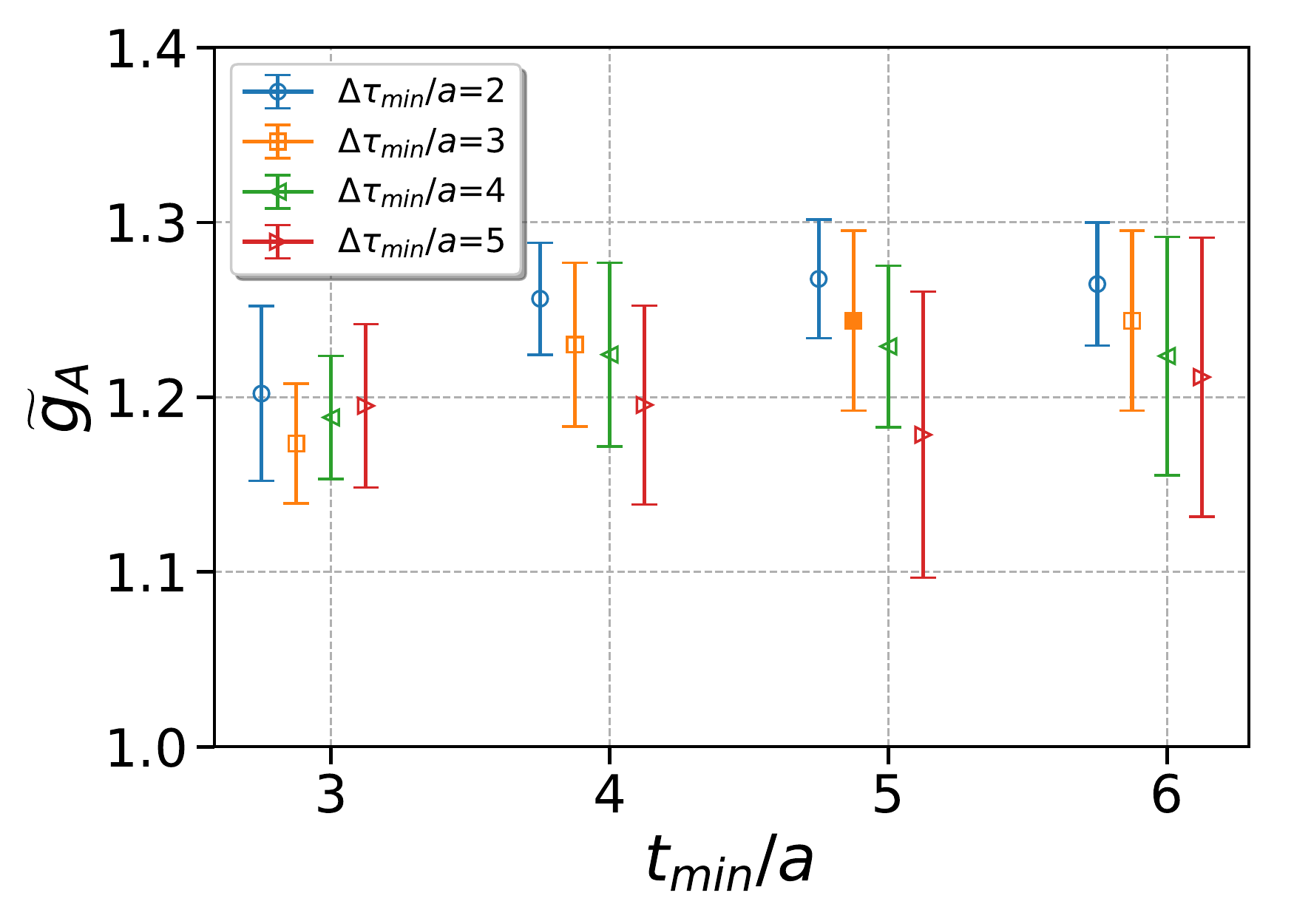}
    \caption{(Color online) The stability plot for the bare vector charge, $\widetilde{g}_V$ (top), and bare axial charge,
        $\widetilde{g}_A$ (bottom), as a function of $t_\text{min}$ and $\Delta\tau_\text{min}$.
        See the caption of Fig.~\ref{fig:stab_M} for further details.}
    \label{fig:stab}
\end{figure}

We plot the stability of the extracted $M_N$ ($\widetilde{g}_V$ and $\widetilde{g}_A$) as a function of $t_\text{min}$ and
$\Delta\tau_\text{min}$ in Fig.~\ref{fig:stab_M} (Fig.~\ref{fig:stab}).
The $x$-axes are different choices of $t_\text{min}$, and the $y$-axes are the corresponding observables.
The four different choices of $\Delta \tau_\text{min}$ are also shown slightly displaced for each~$t_\text{min}$.
The solid squares are the nominal fits with parameters given in Table \ref{tab:param}.
As can be seen in Fig.~\ref{fig:stab_M}, the extracted nucleon mass is stable as a function of $t_\text{min}/a$ and
$\Delta\tau_\text{min}/a$, which illustrates the lack of excited-state contamination in these posteriors.
Similar behavior is seen for $g_V$ in Fig.~\ref{fig:stab}.
The only noticeable structure in the stability plots is for $g_A$, where the observable is stable for $t_\text{min}/a \ge 4$.
Note that as we increase $t_\text{min}/a$ or $\Delta \tau_\text{min}/a$, fewer data are available to fit, and, consequently, the
results become less precise.
Thus, as in Ref.~\cite{Lin:2019pia}, we have demonstrated control over excited-state contamination when extracting matrix elements
from staggered-baryon correlators.

\section{Results}
\label{sec:results}

In this section, we present our Bayesian fitting results and our final renormalized values for the nucleon charges $g_V$ and~$g_A$.
All fitting errors are estimated from $1000$ bootstrap samples.
We take correlations into account by using the same bootstrap samples for both $\widetilde{g}_V$ and $\widetilde{g}_A$.

\subsection{Nucleon Mass}

\begin{figure}
	\centering
	\includegraphics[width=\columnwidth]{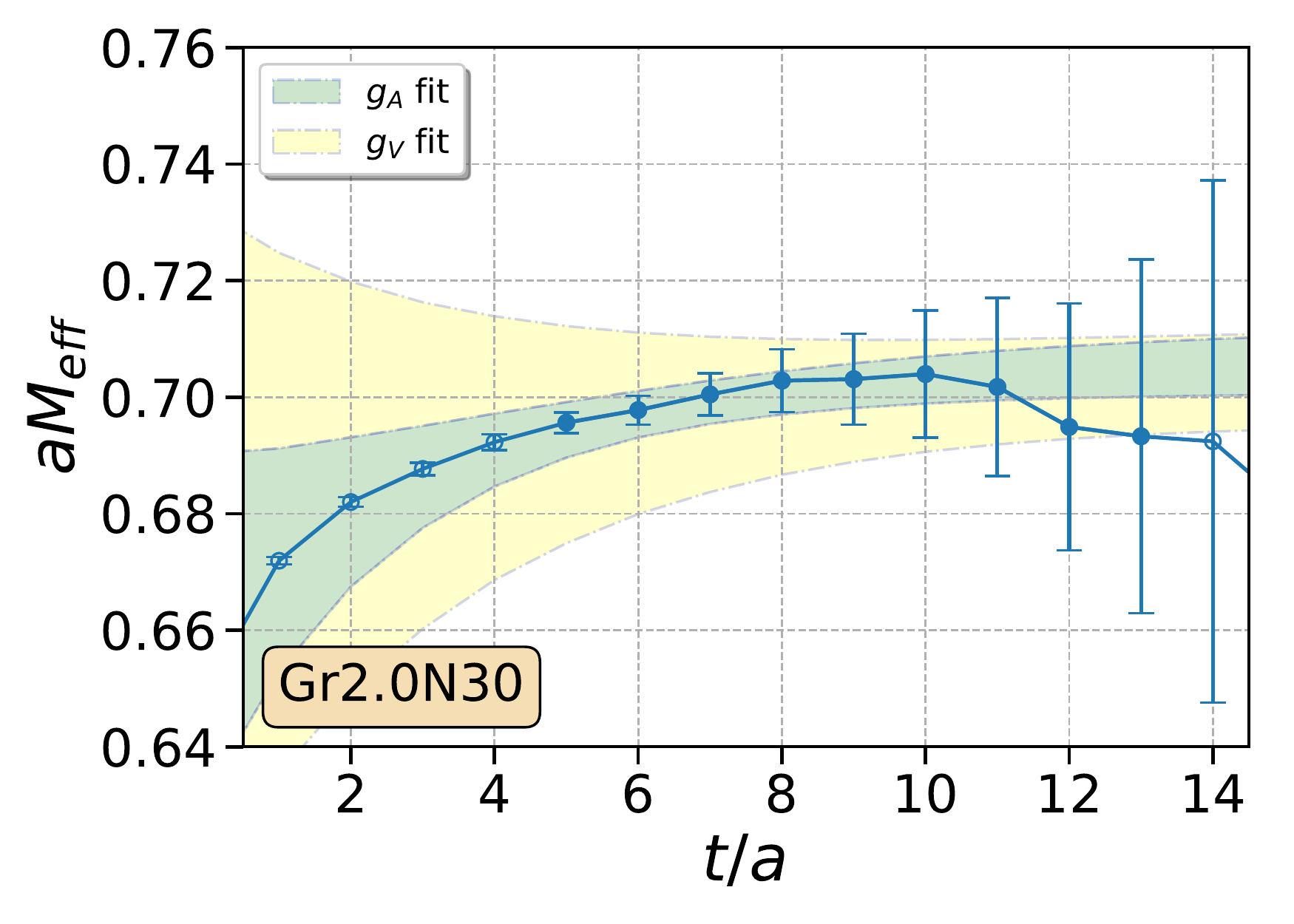}
	\includegraphics[width=\columnwidth]{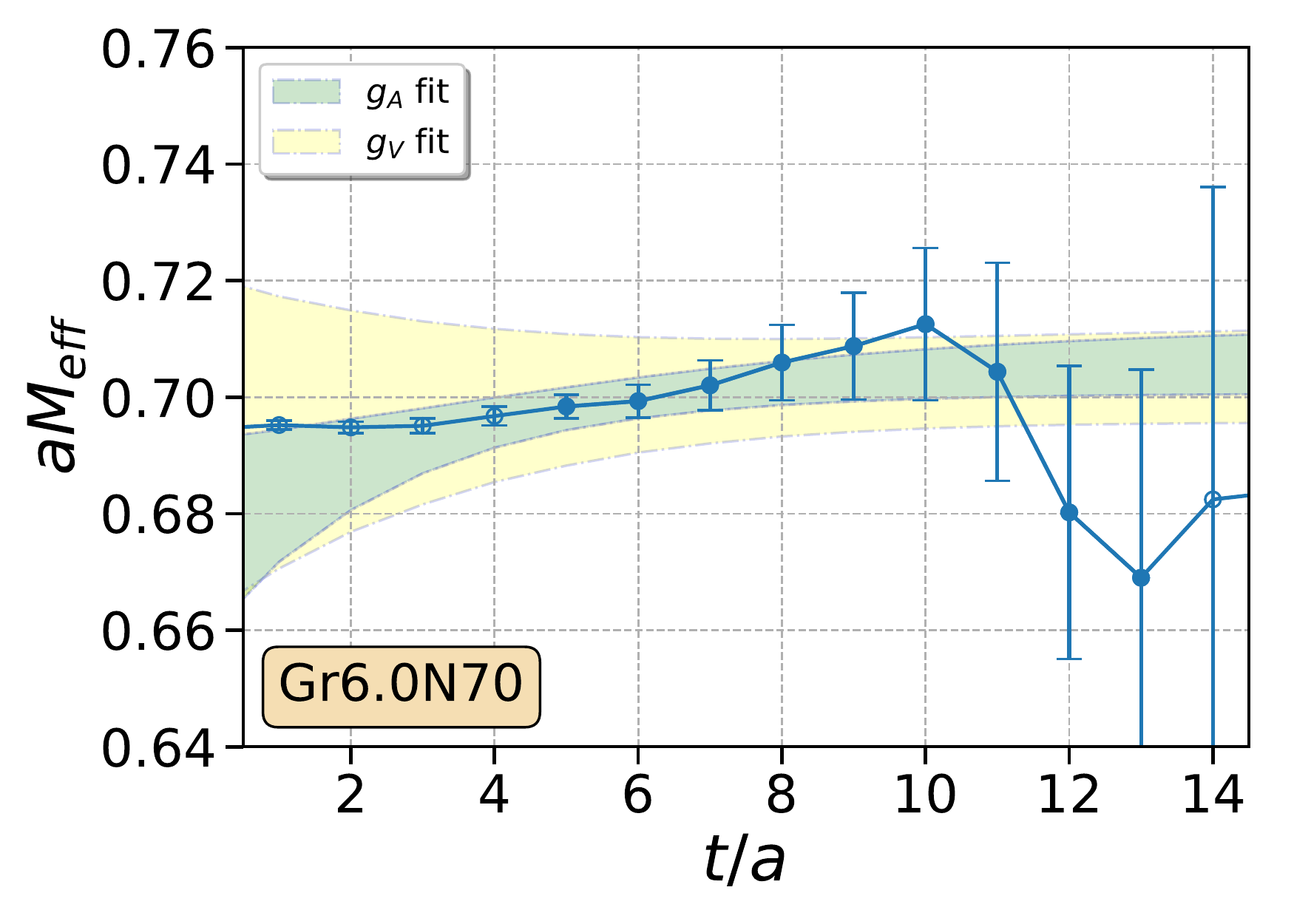}
	\caption{(Color online) %
        Nominal fit results for the effective masses of the optimized correlators as a function of source-sink separation time~$t$.
        The open circles are excluded from the fits.
        Correlators are labeled by their Wuppertal smearing parameters, with RMS radii of $0.2$~fm (Gr2.0N30) or $0.6$~fm
        (Gr6.0N70).
        We perform simultaneous fits to two-point correlators with either the $g_A$ or the $g_V$ three-point correlators.
        Both sets of Wuppertal smeared correlators are included in each fit.
        The green and yellow shading shows the $1\sigma$ bands from fits with either $g_A$ or $g_V$, respectively.}
	\label{fig:2ptfit}
\end{figure}

In Fig.~\ref{fig:2ptfit}, we plot the extracted posterior fitted value for the nucleon mass from simultaneous fits of both smearings
of the optimized two-point correlator and three-point correlator of a given current.
We also plot the nucleon-optimized effective masses.
This effective-mass data is identical to the $t_0/a=6$ data shown in Fig.~\ref{fig:2pt_vary_t0}.
The green-shaded bands are the posterior estimates with the $g_A$ three-point correlators, while the yellow-shaded bands are with
the $g_V$ three-point correlators.
We obtain $aM_N=0.707(6)$ from the $g_A$ fit, and $aM_N = 0.704(9)$ from the $g_V$ fit.

There are some notable features in our fits.
First, the $g_V$ fit has larger posterior uncertainties than the $g_A$ fit.
Both fits include the same information from the two-point correlators, so the difference must arise from the three-point
correlators.
As one can see in Fig.~\ref{fig:gvvarysmear}, the $g_V$ three-point correlators are less sensitive to the Wuppertal smearing than
the $g_A$ correlators.
On the other hand, the $g_V$ three-to-two-point correlator ratios show remarkably little curvature, even at the early times.
This behavior implies that the vector three-point correlators become quickly saturated by the ground state, and therefore provide
limited additional information about the overlap factors and masses than what is contained in the two-point correlators.
The $g_A$ data does not share these features, and thus contains additional information about the two-point posteriors.
This explains why the $g_V$ fit has a less precise nucleon mass than the $g_A$ fit.

For these reasons, we quote the posterior nucleon mass from the $g_A$ fits as the nominal result, which has value
\begin{equation}
    aM_N = 0.707(6), \quad
     M_N = 1141(10)~\text{MeV},
\end{equation}
where the error shown is statistical only.
It is crucial to bear in mind that this result is for a lattice spacing of $a=0.1222(3)$~fm and pion mass of
$M_\pi=305$~MeV~\cite{Bazavov:2017lyh}.
For comparison, a fit including only the two-point correlators yields $aM_N=0.704(9)$, which is identical to the posterior of the
fit with~$g_V$.

In Ref.~\cite{Lin:2019pia}, we computed the nucleon mass at the same lattice spacing but with a physical pion mass, obtaining
$M_N=960(9)~\text{MeV}$.
The difference between these two masses is $\Delta M_N=181(13)~\text{MeV}$, assuming uncorrelated statistical errors.
Given that the pion mass difference between these two ensembles is about $170~\text{MeV}$, $\Delta M_N$ agrees within $1\sigma$ with
the empirical observation that $M_N=800~\text{MeV}+M_\pi$ within a few per~cent~\cite{Walker-Loud:2014iea}.

\subsection{Nucleon \boldmath$g_V$ and $g_A$ charges}
\label{sec:gAgVcharges}

\begin{figure}
	\centering
	\includegraphics[width=0.98\columnwidth]{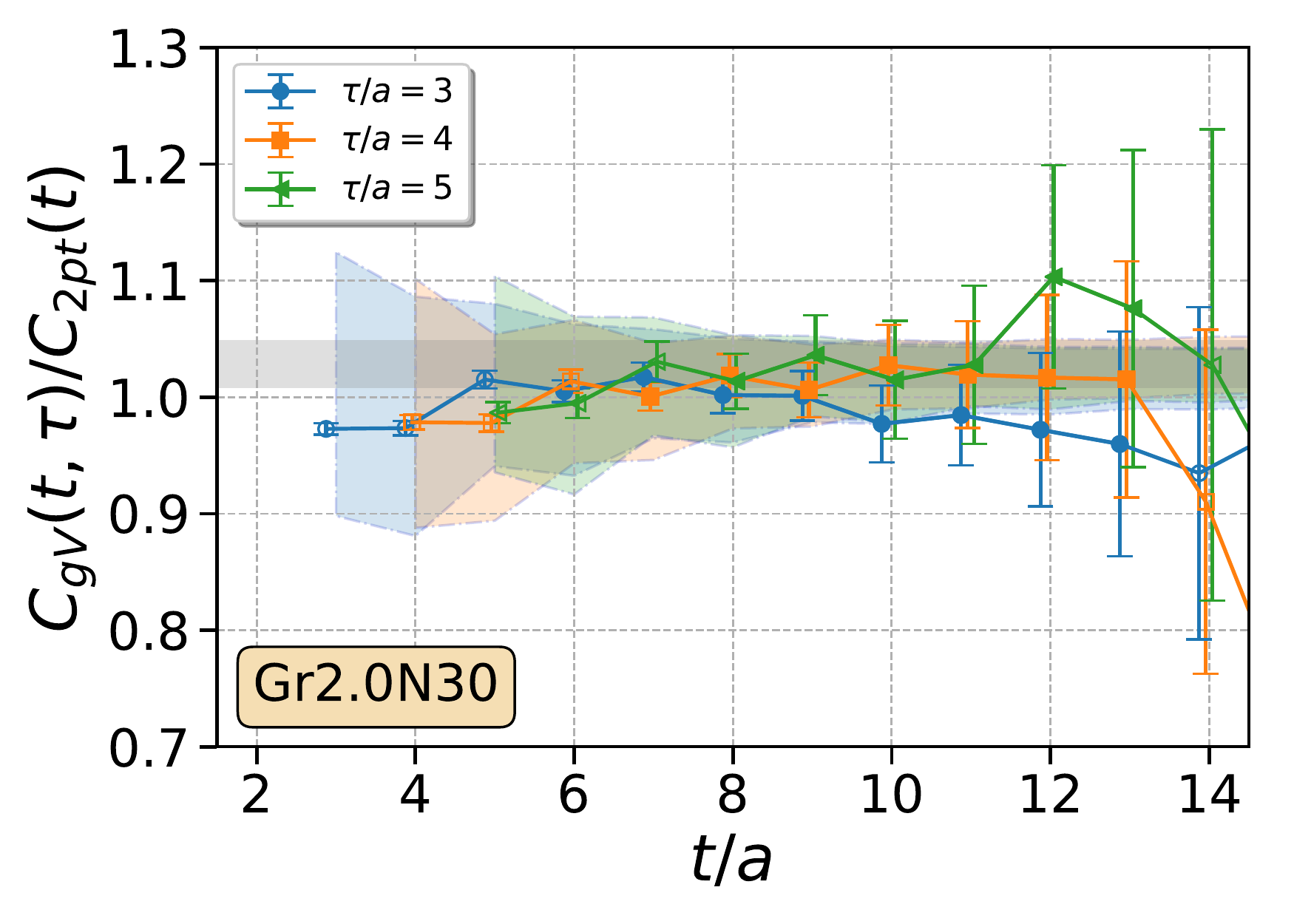}
	\includegraphics[width=0.98\columnwidth]{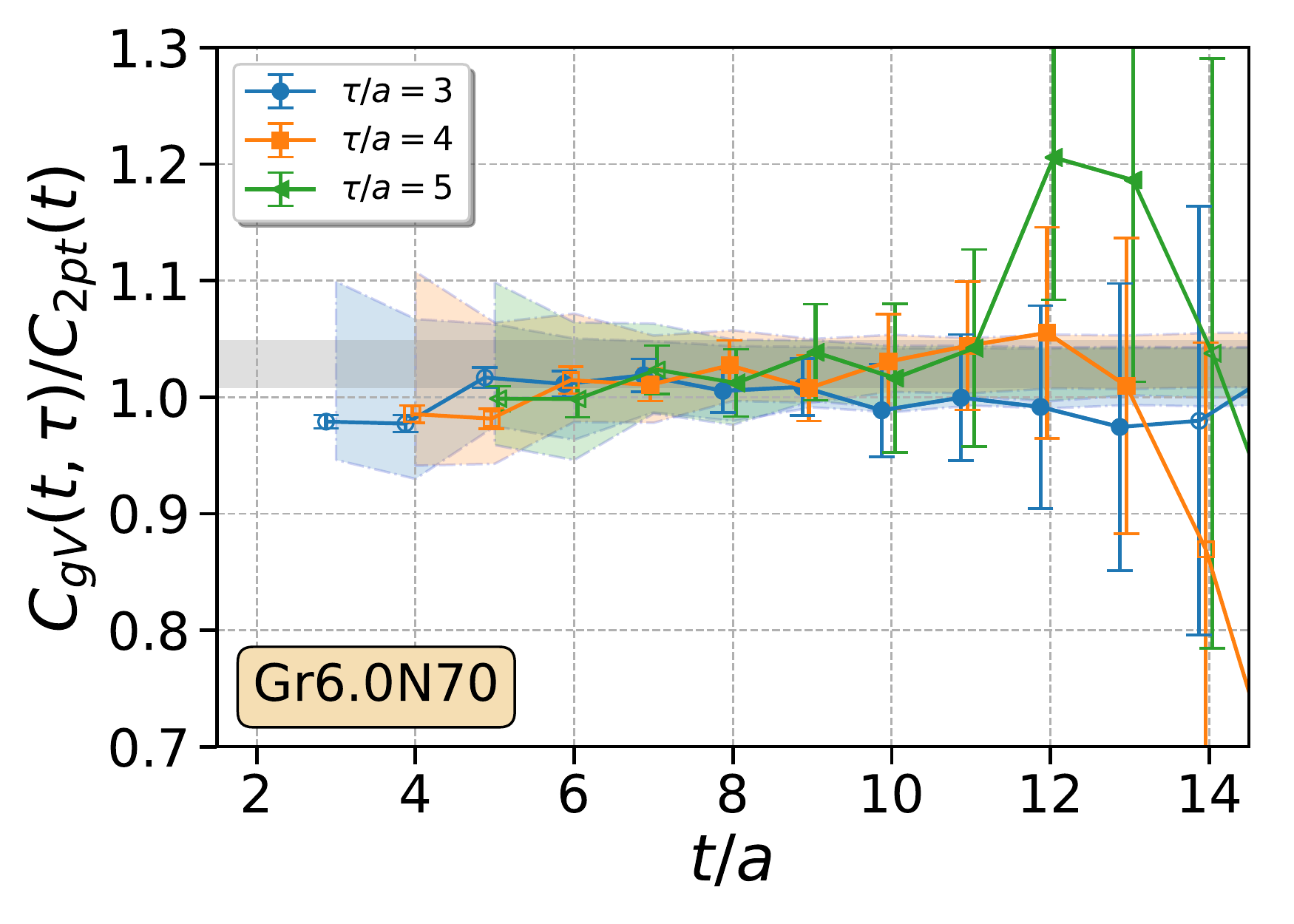}
	\caption{(Color online) %
        Nominal fit results for the optimized three-to-two-point correlator ratio as a function of source-sink separation time~$t$
        and current insertion time $\tau$.
        In the limits $\tau, t-\tau\to\infty$, the optimized three-to-two-point correlator ratios converge to the bare axial charge
        $\widetilde{g}_V$.
        Data points from different current insertion times, $\tau$, are slightly displaced for clarity.
        The filled data points are included in the nominal fit.
        Correlators are labeled by their Wuppertal smearing parameters with rms radii of $0.2$ fm (Gr2.0N30) or $0.6$ fm (Gr6.0N70).
        The $1\sigma$ error bands for the different $\tau$'s are shown in blue, orange, and green, and the $1\sigma$ error band for
        the $\widetilde{g}_V$ posterior is shown in gray.}
	\label{fig:gVfit}
\end{figure}
\begin{figure}
    \centering
    \includegraphics[width=0.98\columnwidth]{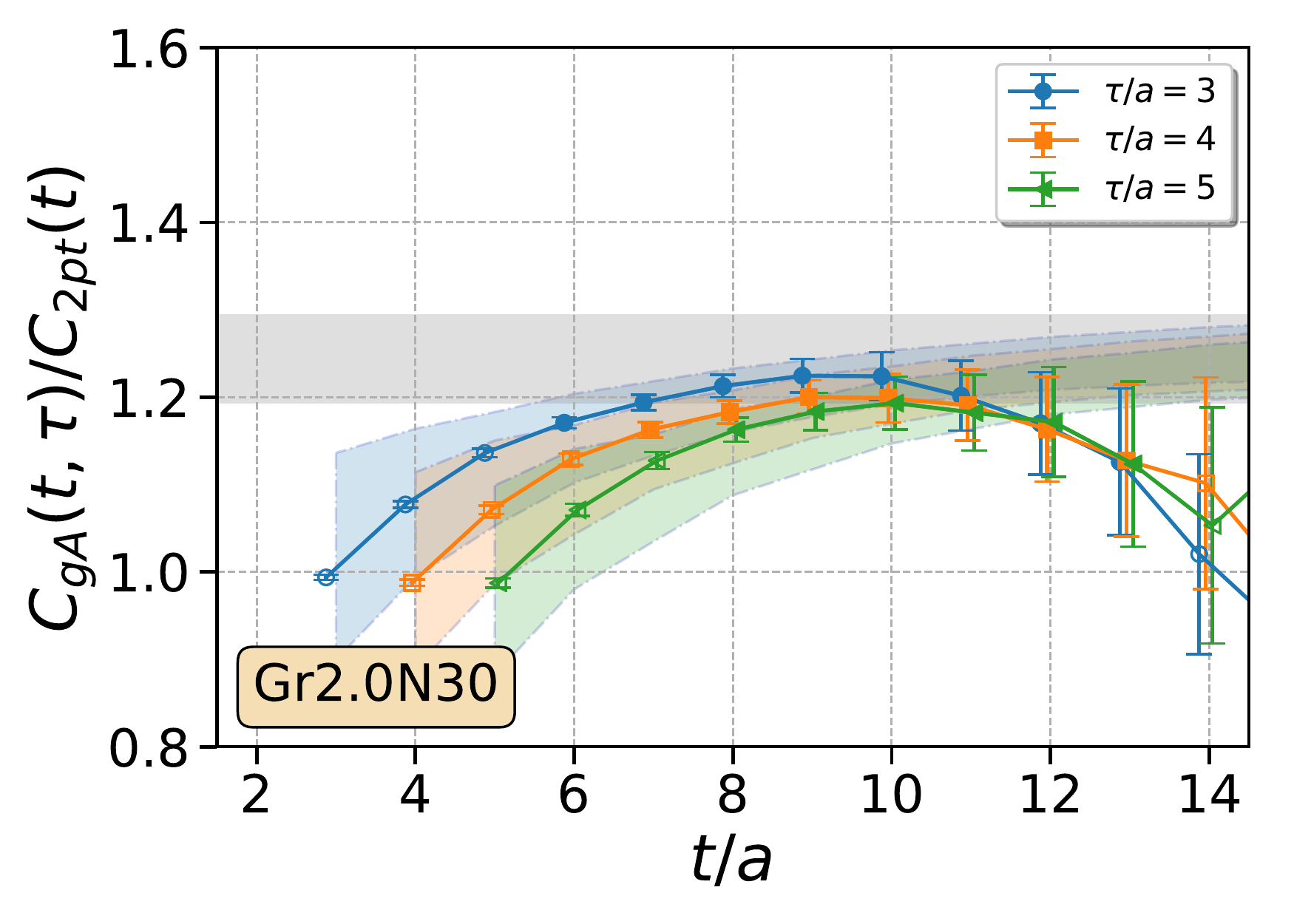}
    \includegraphics[width=0.98\columnwidth]{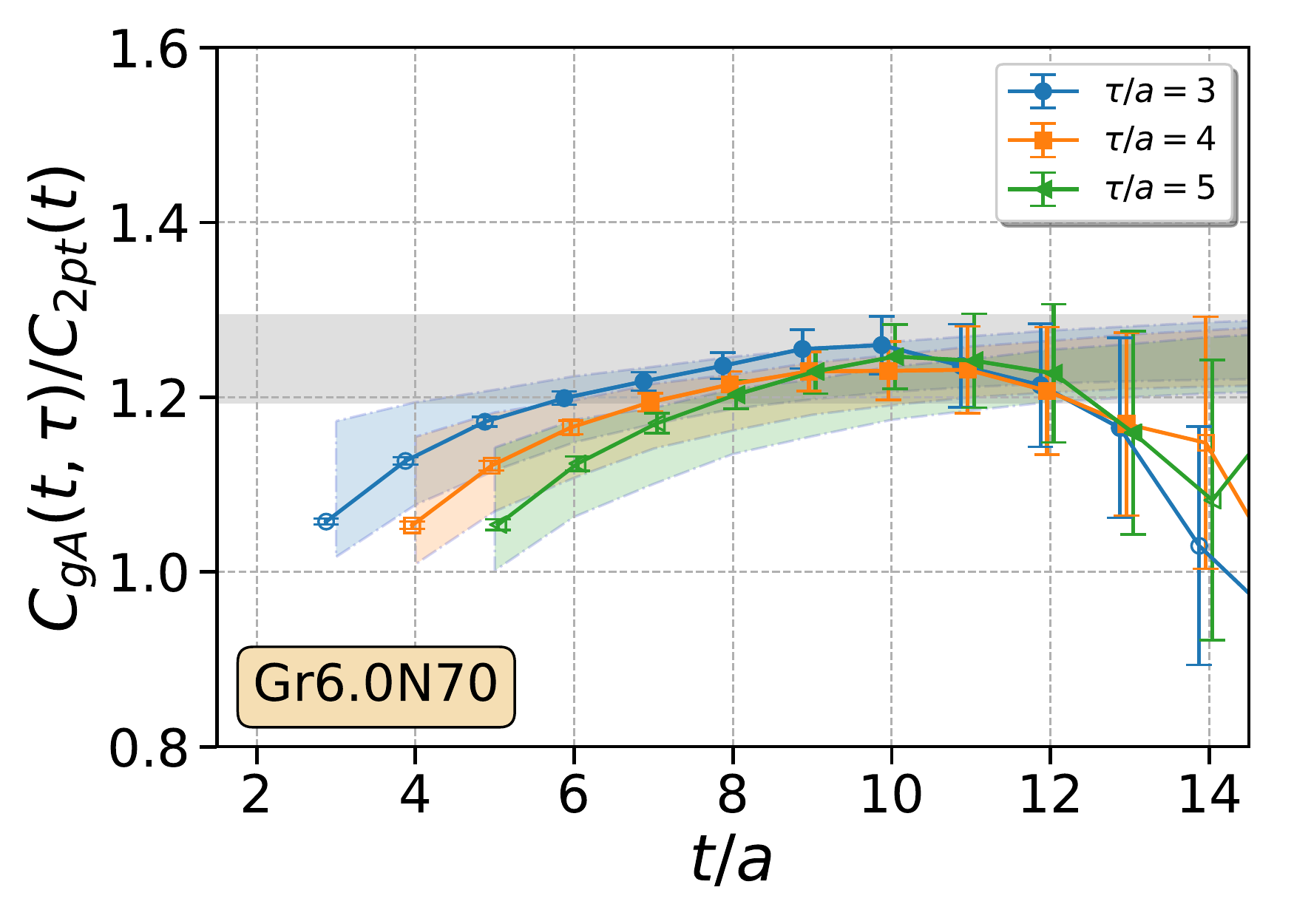}
    \caption{(Color online) %
        Similar to Fig.~\ref{fig:gVfit}, but for the axial-vector three-point correlators $\widetilde{g}_A$.}
    \label{fig:gAfit}
\end{figure}

In Figs.~\ref{fig:gVfit} and~\ref{fig:gAfit}, we plot the optimized $g_V$ and $g_A$ three-to-two-point correlator ratios as a
function of source-sink separation~$t$.
The raw data are identical to the right-hand plots of Figs.~\ref{fig:gvvarysmear} and~\ref{fig:gavarysmear}.
The posterior fit results are superimposed as gray bands.
In the limits $\tau, t-\tau\to \infty$, the data points are seen to converge to these posteriors.
It should be emphasized, however, that the ratio data points are shown only for illustration: we perform direct fits to the
optimized correlators, as discussed in Sec.~\ref{sec:fitting}, in order to obtain results, namely
\begin{align}
    \widetilde{g}_V &= 1.03(2) , \label{eq:baregV} \\
    \widetilde{g}_A &= 1.24(5) . \label{eq:baregA}
\end{align}

It should be mentioned that the $g_V$ and $g_A$ fits have some different features.
First, the residual oscillations from the parity partner matrix element are noticeable in the $g_V$ fit.
Second, the $g_V$ data turn out to be relatively insensitive to the Wuppertal smearing radius.
Both of these features can be observed in Fig.~\ref{fig:gVfit}.
This highlights that there is less uncorrelated data available with which to extract $g_V$ as compared to $g_A$.
In contrast, we observe that the vector correlators in Fig.~\ref{fig:gVfit} contain less positive parity excited state contamination
at early times than the axial-vector correlators in Fig.~\ref{fig:gAfit}.
As such, since the oscillations turn out to be easier to constrain and there is less contribution from the same parity excited
states, we obtain a more precise estimate for $\widetilde{g}_V$ than for $\widetilde{g}_A$.

As discussed in Sec.~\ref{sec:ME}, the remnant chiral symmetry enforces $Z_A=Z_V+\order(am_q)^2$.
Therefore, the ratio of bare charges is renormalized, and we obtain a value of
\begin{align}
    \frac{g_A}{g_V} = \frac{\widetilde{g}_A}{\widetilde{g}_V} = 1.21(5).
    \label{eq:gagvfinal}
\end{align}
where the correlation between $\widetilde{g}_A$ and $\widetilde{g}_V$ is taken into account via bootstrapping.
We can also obtain $Z_V$ by imposing current conversation on a pseudoscalar meson vector-current matrix
element~\cite{Gelzer:2019zwx}.
Then the renormalized charges are
\begin{align}
    g_V &= Z_V\widetilde{g}_V = 1.02(2),\\
    g_A &= Z_A\widetilde{g}_A =  1.23(5), \label{eq:gAfinal}
\end{align}
based on $Z_V=Z_A=0.991(1)$~\cite{Gelzer:2019zwx}.

\section{Discussion and Conclusions}
\label{sec:conclusions}

We have presented two key results in this work.
First, we have shown how to analytically relate the staggered nucleon-like matrix elements with non-trivial tastes to the physical
nucleon matrix elements.
This step is crucial for our on-going program of extracting high-precision nucleon results from staggered fermions.
The general procedure, which can be applied to any staggered baryon matrix element, is outlined in Appendices~\ref{app:matrix}
and~\ref{app:WET}.
Specifically, for the nucleon charges $g_V$ and $g_A$, we summarize our key results for the zero-momentum isovector (axial) vector
three-point correlators in Eqs.~(\ref{eq:3ptraw_gV}) and~(\ref{eq:3ptraw_gA}).
These equations explicitly show the non-trivial normalizations needed to relate the nucleon-like matrix elements to their physical
counterparts.
Our successful computation of $g_V$ and $g_A$ shows that continued use of the 16~irrep of the staggered symmetry group~GTS is
feasible, which is convenient because the~16 contains a single nucleon-like taste in the spectrum~\cite{Lin:2019pia}.

This finding is encouraging, because the additional complexity of staggered baryons, compared with staggered mesons is probably the
reason staggered-baryon matrix elements have not been explored until now.
There are as many meson tastes (16) as bosonic irreps of GTS.
As such, each staggered meson interpolating operator excites only a single taste of meson.
In contrast, there are $64=4^3$ different tastes of a staggered baryon, yet only three unique irreps of GTS, denoted 8, $8'$, and~16
after their dimension.
Consequently, there are not enough unique components of these irreps to accommodate all 64 tastes of baryons, and more than one
taste of the same baryon can appear in each irrep's tower of states.
Choosing an irrep with only one nucleon taste simplifies the correlator analysis and, as we have shown in this paper, allows
for accurate and precise results for nucleon matrix elements.

The second key result of this work is demonstrating the practicality of staggered baryons by computing the isovector nucleon
vector and axial-vector charges.
For this purpose, we choose a single ensemble with $a\approx 0.12$~fm, 2+1+1 flavors in the sea, and, when using identical sea and
valence HISQ quarks, $M_{\pi}=305$~MeV.
With approximately 7000 measurements and techniques designed to handle staggered correlators, we find few-percent statistical
uncertainty.
Our final values for $g_V$, $g_A$, and $g_A/g_V$ on this ensemble are
\begin{align}
    g_V             &= 1.02(2), \\
    g_A             &= 1.23(5), \\
    \frac{g_A}{g_V} &= 1.21(5).
\end{align}
The conservation of the vector charge, $g_V=1$, is a non-trivial verification of our methodology.

As discussed in Sec.~\ref{sec:fitting}, we include two positive-parity states and one negative-parity state in our fit function.
The number of matrix elements included in the fit grows quadratically as a function of the states included.
With more precise data, we could constrain more matrix elements.
Alternatively, we could also impose tighter priors on the transition matrix elements and overlap factors, for example with the
empirical Bayes method~\cite{Gupta:2018qil}.
This proof-of-concept study does not attempt a full calculation with all errors included, so we leave exploration of those options
for future work.

The same ensemble has been used by both the CalLat~\cite{Berkowitz:2017gql,*Chang:2018uxx} and PNDME~\cite{Gupta:2018qil}
collaborations in their calculations of $g_V$ and~$g_A$.
CalLat uses M\"obius domain-wall fermions for the valence quarks, while PNDME uses Wilson fermions with the clover action.
CalLat defines $Z_V$ by demanding $g_V=Z_V\widetilde{g}_V =1$ and uses the remnant chiral symmetry to set $Z_A=Z_V$.
They then quote $\widetilde{g}_V=1.021(2)$ and $g_A=1.21(1)$.
PNDME determines $Z_V$ and $Z_A$ independently via the regularization-independent symmetric momentum-subtraction scheme, commonly
known as RI-sMOM, and quote $g_V=0.97(2)$, $g_A=1.21(4)$, and $g_A/g_V=1.25(2)$.
Our result is consistent with both, despite the different choices of valence-quark formulation.
Other calculations in the literature, which have $2+1$ flavors in the sea, have not been performed at values of $a$, $M_{\pi}$, and
physical volume close enough to ours to allow a straightforward comparison.

With an eye towards sub-percent determinations of the axial charge, it is instructive to compare how the precision on~$g_A$ is
influenced by each collaboration's data and methodology.
Presumably influenced by the common ensemble, the three analyses share a few common aspects.
First is the use of eight sources (with high-precision solutions of the Dirac equation) per gauge-field configuration, so the
raw statistics are about the same.
Second, the time range of the central fits for the two-point correlators turns out to be the same: $t_\text{max}+1-t_\text{min}=8$.
Third, all three collaborations simultaneously fit a correlator containing the matrix element with the two-point 
correlators.
Last, PNDME and we use time ranges in the central fits of the three-point correlators, such that there are 21 data points in the
fit.

In addition, each collaboration employs techniques to improve the signal.
We have two smeared sinks and start with $4\times4$ matrix correlators, which is natural and necessary with our choice of staggered
irrep.
We apply the GEVP to the $4\times4$ matrix for each smearing radius to find the optimal source and sink operators for the
nucleon.
PNDME increases statistics via the truncated-solver method with bias correction~\cite{Bali:2009hu,Blum:2012uh}.
CalLat reduces noise with an $a$-independent number of steps of a gradient flow~\cite{Luscher:2010iy}.
In the future, we could easily take advantage of the truncated-solver method, while the gradient flow would prevent us from using
numerous technical results from the Fermilab Lattice and MILC collaborations, such as lattice-spacing and renormalization-factor
determinations.

A more striking difference is CalLat's introduction of the currents into a propagator in a way inspired by the Feynman-Hellmann
theorem~\cite{Chang:2017oll}.
A key feature of the technique is that instead of a three-point function, the matrix element lies within another two-point function.
Thus, the CalLat method requires a fit to a single time variable instead of two; indeed the matrix element pops out of a fit to the
ratio of the two two-point correlators.

In the end, the relative precision on $g_A$ is quoted as 1\%, 3\%, and 4\%~for CalLat~\cite{Berkowitz:2017gql,*Chang:2018uxx},
PNDME~\cite{Gupta:2018qil}, and this work, respectively.
One should bear in mind, however, the effective number of components per site, which are four for Wilson fermions, eight for
staggered fermions (corresponding to the corners of the unit cube), and $4L_5$ for domain-wall fermions (where $L_5$ is the extent
of the fifth dimension; $L_5=8$ in Ref.~\cite{Berkowitz:2017gql,*Chang:2018uxx}).
Taking the number of components into account but ignoring algorithmic speed-ups from the code or specific features of each action,
the cost for given precision is roughly the same.
It would, therefore, be interesting to explore the truncated-solver and Feynman-Hellmann-inspired methods with staggered fermions.

This work sets the foundation needed to continue a program of precise nucleon form-factor calculations.
Calculations of the vector and axial-vector form factors at nonzero momentum transfer are indeed underway on the same ensemble
as used here.
Further we, have started computing $g_V$ and $g_A$ on the same ensembles used in Ref.~\cite{Lin:2019pia}.
These ensembles have physical pion masses, and a range of lattice spacings to enable a continuum extrapolation.

\appendix
\section{Relating Staggered-QCD Matrix Elements to QCD Matrix Elements}
\label{app:matrix}

Lattice gauge theory with staggered fermions can be thought of as an extension of QCD with four degenerate flavors, called tastes,
for each quark.
The associated taste symmetry allows for many more composite states which can have non-trivial taste structures.
We call states that have non-trivial taste ``baryon-like'' states, to distinguish from the physical single-taste baryons.
In this work, we focus on the nucleon and restrict ourselves to that case going forward.
The nucleon-like states can be mapped onto the physical nucleon states through appropriate flavor-taste symmetry transformations.
This allows the freedom to choose which nucleon-like state to study in order to extract observables.
As highlighted in Ref.~\cite{Lin:2019pia}, the two-point correlator data constructed from nucleon-like states are easier to analyze
than their physical counterparts due to the smaller multiplicity of tastes in the spectrum.
However, one needs the mapping from the specific nucleon-like state to the physical state.

We use isospin-$\frac{3}{2}$, GTS-16 nucleon-like interpolating operators to extract nucleon observables, since the spectrum
contains only a single nucleon-like state.
The relationship between the nucleon-like matrix elements and the physical nucleon matrix elements is, unfortunately, not at all
transparent.
In this and the following appendices, we will establish the relationship between the $16$-irrep nucleon-like matrix elements and the
single-taste physical nucleon matrix elements.

Bailey~\cite{Bailey:2006zn} inferred the spectrum of staggered baryons by subducing nucleon-like representations of the full
$\text{SU}(8)_{FT}$ flavor-taste symmetry of the continuum limit into~GTS.
We expand that work to matrix elements.
Specifically, we will demonstrate how one can apply the generalized Wigner-Eckart theorem to $\text{SU}(4)$ and relate the lattice
nucleon-like matrix elements to the physical tasteless nucleon matrix elements through appropriate normalization factors, which are
generalized Clebsch-Gordan coefficients.
The procedure outlined here can be applied to any staggered baryon matrix elements in any $\text{SU}(n_f)\times\text{GTS}$
flavor-taste~irrep.

Following the notation from Ref.~\cite{Lin:2019pia}, we first determine the continuum quantum numbers of the nucleon-like states
that subduce into the $16$ irrep of GTS.
This step is needed for the generalized Wigner-Eckart theorem.
We focus on the continuum symmetry group $\text{SU}(2)_S \times \text{SU}(8)_{FT}$, where $\text{SU}(2)_S$ is the spin symmetry
and $\text{SU}(8)_{FT}$ is the flavor ($F$) and taste ($T$) symmetry for two equal-mass flavors.
This group breaks on a discrete lattice to the unbroken flavor symmetry subgroup $\text{SU}(2)_F$ and the ``geometric
timeslice group'' (GTS) \cite{Golterman:1984dn,Bailey:2006zn}.
GTS can be decomposed into~\cite{Lin:2019pia,Meyer:2017ddy}
\begin{equation}
	\text{GTS} = \left((\mathds{Q}_8 \rtimes \text{SW}_3) \times \text{D}_4\right)/\mathds{Z}_2,
	\label{eq:gtsdecomp}
\end{equation}
where $\mathds{Q}_8$ is generated by the discrete taste transformations $\{\Xi_{12}, \Xi_{23}\}$, $\text{SW}_3$ by the cubic
rotations $\{R_{12}, R_{23}\}$, and $\text{D}_4$ by the discrete taste and spatial inversion transformations $\{\Xi_{123}, I_S\}$.
(These symbols are all defined in the appendix of Ref.~\cite{Lin:2019pia}.) %

\begin{widetext}
The subgroup chain we work with is%
\footnote{Various $\mathds{Z}_N$ quotient factors are often omitted for clarity.
They are only necessary to avoid overcounting group elements (for example,
$\text{SU}(4)_T\supset(\text{SU}(2)_{\mathds{Q}_8}\times\text{SU}(2)_{\text{D}_4})/\mathds{Z}_2$).}
\begin{align}
	\text{SU(2)}_S \times \text{SU}(8)_{FT} \times P & \supset 
        \text{SU(2)}_S \times \text{SU}(2)_{F} \times \text{SU}(4)_T  \times P \nonumber \\
    & \supset \text{SU}(2)_S \times \text{SU(2)}_F \times \text{SU}(2)_{\mathds{Q}_8} 
        \times \text{SU}(2)_{\text{D}_4} \times \text{U}(1)_{\text{D}_4}  \times P \nonumber \\
    & \supset \text{SU}(2)_F \times \text{GTS} \times P,
    \label{eq:subduce}
\end{align}
where $P = I_S\Xi_4$ becomes the usual parity operation in the continuum limit~\cite{Golterman:1984dn}.
The factor $\text{SU}(2)_{\text{D}_4}$ on the second line arises from decomposing the $\text{SU}(4)_T$ taste symmetry onto a
discrete lattice, which leads to the factor $\text{D}_4$ in Eq.~(\ref{eq:gtsdecomp}), combined with the $\text{U}(1)_{\text{D}_4}$
phase factor.
Note that in Ref.~\cite{Lin:2019pia} we omitted the $\text{U}(1)_{\text{D}_4}$ factor, but here we make it explicit.
The other the groups are defined and explained in Ref.~\cite{Lin:2019pia}.
\end{widetext}

\subsection{Using shift symmetries to relate staggered correlators}
\label{app:nucleon_on_lattice}

The goal is to assign continuum quantum numbers of $\text{SU}(2)_S\times \text{SU}(2)_F\times\text{SU}(4)_T$ to each nucleon-like
state created by every component of the $16$ irrep.
We begin by investigating the continuum quantum numbers of the simplest nucleon-like states created by the $16$ irrep.
Afterwards, we can use the lattice symmetry transformations to obtain the remaining components.

We can form nonvanishing two-point correlation functions is by contracting any one of the 16 irrep components with the same
component on a later timeslice.
One can then apply lattice rotations and shifts%
\footnote{We use the convention of staggered phases, $\eta_1(x)=(-1)^{x_4}$, $\eta_2(x)=(-1)^{x_4+x_1}$,
$\eta_3(x)=(-1)^{x_4+x_1+x_2}$, and $\eta_4(x)=1$~\cite{Bernard:1997mz}, which affects the phases
appearing in the lattice rotations and shifts.} %
to show that these 16 two-point correlators are identical in the ensemble average.

The $16$ irrep components split into two sets of 8 different components that reside on the eight corners of a cube (see the appendix
of Ref.~\cite{Lin:2019pia} for explicit constructions).
The construction of nonvanishing three-point correlator data also depends on the current insertion.
For the local vector and axial-vector currents we use in this work, the zero-momentum three-point correlators do not vanish if and
only if the source and sink interpolators are identical.
Correlators constructed from the same set of 8 components can be related to each other with the lattice shift symmetries.

To summarize, this means that the nonvanishing two-point correlators satisfy
\begin{align}
    \sum_{\vec{x}} \left\langle  B^{16}_{+\vec{M}}(\vec{x},t)\overline{B}^{16}_{+\vec{M}}(0)\right\rangle =
    \sum_{\vec{x}} \left\langle  B^{16}_{s\vec{N}}(\vec{x},t)\overline{B}^{16}_{s\vec{N}}(0)\right\rangle 
    \label{eq:2ptrelation}
\end{align}
where the superscript denotes the $16$ irrep operators, $\vec{M}$ and $\vec{N}$ are equal to any one of the eight corners of the
cube, and $s=\pm 1$ are the eigenvalues of the lattice rotation $R_{12}$ for $\vec{M} = \vec{N} = \vec{0}$.
The notation is defined in detail in Ref.~\cite{Lin:2019pia}.

We are using local currents $\Jlat=\Vlat,\Alat$ in this work, so the nonvanishing three-point correlators satisfy
\begin{align}
	&\sum_{\vec{x},\vec{y}} \left\langle  B^{16}_{s\vec{M}}(\vec{x},t)\Jlat(\vec{y},\tau)
        \overline{B}^{16}_{s\vec{M}}(0)\right\rangle = 
    \label{eq:threepointcorrelator} \\
	&~~~~~~~~~~~\mathcal{S}_\Jlat(\vec{N}-\vec{M})\sum_{\vec{x},\vec{y}}
        \left\langle B^{16}_{s\vec{N}}(\vec{x},t)\Jlat(\vec{y},\tau)\overline{B}^{16}_{s\vec{N}}(0)\right\rangle ,
    \nonumber
\end{align}
where $\mathcal{S}_J(\vec{A})=\pm 1$ is a sign factor that depends on both $\Jlat$ and $\vec{A}$.
Its specific value can be determined by applying a lattice shift symmetry transformation between $\vec{M} $ and $\vec{N}$.
For the currents used in this work, it is identical to the sign factor appearing in the construction of the staggered
current~$\Jlat$.
For example, $\mathcal{S}_{V}(\vec{A})=(-1)^{(A_x+A_y+A_z)/a}$ for the $\gamma_4\otimes\xi_4$ vector current and
$\mathcal{S}_{A}(\vec{A})=(-1)^{A_z/a}$ for the $\gamma_z\gamma_5\otimes\xi_z\xi_5$ axial current.
The currents and phase factors are also defined in Eqs.~(\ref{eq:JV}) and (\ref{eq:JA}).
For a general current (other than the local currents used here), however, it might be necessary to have different interpolating
operators at the source and sink.
In that case, the phase factors in the general version of Eq.~(\ref{eq:threepointcorrelator}) would still be obtained from the
lattice shift symmetries.

Going forward, it is sufficient to study the correlator with component $\vec{N}=0$ located at the origin of the staggered unit cube,
$\sum_{\vec{x}}B^{16}_{\pm\vec{0}}(\vec{x},t)$.
Then, owing to Eq.~(\ref{eq:threepointcorrelator}), the other seven
components follow immediately.

The quantum numbers of the nucleon-like states created by $\sum_{\vec{x}}B^{16}_{\pm\vec{0}}(\vec{x},t)$ will be denoted as
$\big|[\frac{3}{2},\frac{3}{2}]_F[16, \pm\vec{0}]_\text{GTS}\big\rangle$.
The first bracket gives the unbroken $\text{SU}(2)_F$ flavor quantum numbers, which here has total and $z$-component isospins
$\frac{3}{2}$, and the second bracket denotes the $16$ irrep with the eigenvalues of~$R_{12}$.

\subsection{Quantum numbers of nucleon-like states}

Next, we must find a convenient basis for the continuum nucleon-like states and then subduce them down to the
$\big|[\frac{3}{2},\frac{3}{2}]_F[16, \pm\vec{0}]_\text{GTS}\big\rangle$ lattice states.
From Eq.~(\ref{eq:subduce}), we want to track the quantum numbers of
$\text{SU}(2)_S\times\text{SU}(2)_F\times\text{SU}(4)_T$ and may ignore the passive phase $\text{U}(1)_{\text{D}_4}$ and parity
$P=+1$ factors.
From the group subduction presented in Ref.~\cite{Lin:2019pia, Bailey:2006zn}, the $16$ irrep is subduced from the continuum
spin-flavor-taste irrep via
\begin{align}
	&\text{SU}(2)_S\times\text{SU}(2)_F\times\text{SU}(4)_T \supset \nonumber\\
    &~~~~~~~~~\text{SU}(2)_S\times\text{SU}(2)_F\times\text{SU}(2)_{\mathds{Q}_8}\times\text{SU}(2)_{\text{D}_4}\\
	&\left(\frac{1}{2}, \frac{3}{2}, 20_M\right) \to \nonumber \\
	&~~~~~~~~~
	\left(\frac{1}{2}, \frac{3}{2}, \frac{1}{2}, \frac{3}{2}\right) \oplus 
	\left(\frac{1}{2}, \frac{3}{2}, \frac{3}{2}, \frac{1}{2}\right) \oplus
	\left(\frac{1}{2}, \frac{3}{2}, \frac{1}{2}, \frac{1}{2}\right).
	\label{eq:groupcomp}
\end{align}
Here we have adopted a convention that labels non-$\text{SU}(2)$ group irreps by their dimensions and subscript $M$~(mixed),
$S$~(symmetric), or $A$~(antisymmetric).
The irreps of $\text{SU}(2)$ are denoted with standard spin notation.

The task of classifying a general irrep of $\text{SU}(4)$ amounts to finding the maximal set of commuting operators and uniquely
labeling the states by their eigenvalues; for a general $\text{SU}(4)$ irrep, there are 6 eigenvalues to
classify~\cite{Hecht:1969ck}.
Because there are no degenerate irreps when decomposing any of the irreps in this work from $\text{SU}(4)$ into
$\text{SU}(2)\times\text{SU}(2)$, we can use the eigenvalues of the pair of $\text{SU}(2)$ factors to identify $\text{SU}(4)$
states.
Therefore, only 4 of those 6 eigenvalues are necessary to completely characterize the states.
As such, the 4 eigenvalues of each state can be uniquely identified with two pairs of the $|L^2, L_z\rangle$ quantum numbers.

Given Eq.~(\ref{eq:groupcomp}), we notice that
$20_M\to\left(\frac{1}{2},\frac{1}{2}\right)\oplus\left(\frac{3}{2},\frac{1}{2}\right)\oplus\left(\frac{1}{2},\frac{3}{2}\right)$.
We seek to find the four quantum numbers for the states after decomposition of the $20_M$ irrep of $\text{SU}(4)_T$ into the
subgroup $\text{SU}(2)_{\mathds{Q}_8}\times\text{SU}(2)_{\text{D}_4}$.
We write the continuum nucleon-like states kets
\begin{equation} 
	\left|
	\left[\frac{1}{2}, m^S\right]_S
	\left[\frac{3}{2},\frac{3}{2}\right]_F
	\bigg[j^{\mathds{Q}_8},m^{\mathds{Q}_8}\bigg]_{\mathds{Q}_8}
	\bigg[j^{\text{D}_4},m^{\text{D}_4}\bigg]_{\text{D}_4}
	\right\rangle.
	\label{eq:conket}
\end{equation}
Each bracket represents the standard spin quantum numbers of one of the $\text{SU}(2)$ group factors, distinguished by the
superscripts and subscripts: $S$~(spin), $F$~(flavor), $\mathds{Q}_8$ ($\text{SU}(2)_{\mathds{Q}_8}$), and $\text{D}_4$
($\text{SU}(2)_{\text{D}_4}$).
This ket serves as the irrep basis for both $\text{SU}(2)_S\times\text{SU}(2)_F\times\text{SU}(4)_T$ and
$\text{SU}(2)_S\times\text{SU}(2)_F\times\text{SU}(2)_{\mathds{Q}_8}\times\text{SU}(2)_{\text{D}_4}$.

\subsection{Matching the continuum and lattice nucleon-like states}

Now that we have established an appropriate basis for the nucleon-like states, both on the lattice and in the continuum, we are
ready match the two sets.
In particular, we are interested in which linear combination of states from Eq.~(\ref{eq:conket}) combine to subduce into the
lattice states $\big|[\frac{3}{2},\frac{3}{2}]_F[16, \pm\vec{0}]_\text{GTS}\big\rangle$ of interest.
For the $16$ irrep nucleon-like states, we have shown in Ref.~\cite{Lin:2019pia} that $j^{\mathds{Q}_8}=\frac{3}{2}$ and
$j^{\text{D}_4} = \frac{1}{2}$.
Consequently, we only need to determine $m^S$, $m^{\mathds{Q}_8}$, and $m^{\text{D}_4}$.

We start with determining $m^{\text{D}_4}$ of $\text{SU}(2)_{\text{D}_4}$ from Eq.~(\ref{eq:conket}).
To do so, it is illuminating to study the decomposition
\begin{equation}
    \text{SU}(2)_{\text{D}_4}\times \text{U}(1)_{\text{D}_4}\times P  \to \{I_S\},
    \label{eq:IS}
\end{equation}
where $\{I_S\}$ is the group generated by the lattice spatial inversion.
As Eq.~(\ref{eq:IS}) shows, $I_S$ receives contributions from three different factors: the taste factor $\text{SU}(2)_{\text{D}_4}$,
a phase factor $e^{-i\pi/2} = -i$ from $\text{U}(1)_{\text{D}_4}$ to match the eigenvalues of $I_S$, and
the continuum-limit parity $P = I_S\Xi_4$.
For the spin-\onehalf\ irreps of $\text{SU}(2)_{\text{D}_4}$, which include the 16 irrep nucleons \cite{Lin:2019pia}, the matrix
representation of $I_S$ is the tensor product of those three factors
\begin{align}
    e^{i\sigma_3\pi/2}\otimes e^{-i\pi/2} \otimes +1 = 
    \begin{bmatrix}
	    1 & 0 \\
	    0 & -1
    \end{bmatrix} = 
    \sigma_3 = I_S,
    \label{eq:ISMatrix}
\end{align}
where $\sigma_3$ is the third Pauli matrix.
The representation in Eq.~(\ref{eq:ISMatrix}) can be mapped onto the groups in Eq.~(\ref{eq:IS}).
The first factor arises from the $180$ degrees rotation in the ``$x$-$y$ plane'' of the spin-\onehalf\ representation of
$\text{SU}(2)_{\text{D}_4}$, the second $e^{-i\pi/2}$ phase is from $\text{U}(1)_{\text{D}_4}$, and the $+1$ from parity.
As can be seen from Eq.~(\ref{eq:ISMatrix}), for the spin-\onehalf\ irrep of $\text{SU}(2)_{\text{D}_4}$, the $I_S$ matrix admits
$\pm1$ eigenvalues which arise from the $m^{\text{D}_4}=\pm\onehalf$ components of $\sigma_3$.
Since the nucleon is a positive-parity state with $I_S=1$, we assign $m^{\text{D}_4} = \frac{1}{2}$ to the
$\big|[\frac{3}{2},\frac{3}{2}]_F[16, \pm\vec{0}]_\text{GTS}\big\rangle$ lattice states.

We now consider the quantum numbers of $m^S$ and $m^{\mathds{Q}_8}$.
The $16$ irrep components can be labeled by the irreps of $\text{W}_{3}=\text{SW}_3\times\{1,I_S\}$, where $\text{SW}_3$ is the
cubic rotation group, as~\cite{Golterman:1984dn}
\begin{equation}
	16 \to E^+ \oplus E^- \oplus T_1^+ \oplus T_1^- \oplus T_2^+ \oplus T_2^-, 
\end{equation}
where $E$ is the two-dimensional irrep of $\text{SW}_3$, $T_1$ and $T_2$ are the different three-dimensional irreps of
$\text{SW}_3$, and the superscripts show the eigenvalues of~$I_S$.
By applying lattice rotations to $\big|[\frac{3}{2},\frac{3}{2}]_F[16, \pm\vec{0}]_\text{GTS}\big\rangle$, we can show they belong
to the two-dimensional $E^+$ irrep of~$\text{W}_3$.

Subducing $\text{SU}(2)_{\text{SW}_3}\subset\text{SU}(2)_{\mathds{Q}_8}\times\text{SU}(2)_{S}$ to the lattice angular momentum of
$\text{SW}_3$ is a problem common to all fermion formulations~\cite{Basak:2005ir}.
We can write the irrep components of $\text{SU}(2)_{\text{SW}_3}$ that subduce into $E$ as~\cite{Basak:2005ir}
\begin{align}
	\left| \bigg[2, 0 \bigg]_{\text{SW}_3}\bigg[\frac{1}{2},\frac{1}{2}\bigg]_{\text{D}_4}\right\rangle
	&\to 
	\left|\bigg[16, +\vec{0}\bigg]_\text{GTS}\right\rangle \nonumber \\ &
	\to 
    \left|\bigg[E^+,+\bigg]_{\text{W}_3}\right\rangle
\end{align}
\begin{align}
	&\frac{1}{\sqrt{2}}\left(\left| \bigg[2, 2 \bigg]_{\text{SW}_3}\bigg[\frac{1}{2},\frac{1}{2}\bigg]_{\text{D}_4}\right\rangle +
	\left| \bigg[2, -2 \bigg]_{\text{SW}_3}\bigg[\frac{1}{2},\frac{1}{2}\bigg]_{\text{D}_4}\right\rangle\right) \nonumber \\
    & \hspace{3em}
    \to \left|\bigg[16, -\vec{0}\bigg]_\text{GTS}\right\rangle
        \to \left|\bigg[E^+,-\bigg]_{\text{W}_3}\right\rangle
	\label{eq:diagtogts}
\end{align}
where the irreps of $\text{SU}(2)_{\text{SW}_3}$ are again labeled by the total and $z$-component of angular momentum, and the
arrows indicate the subduction from continuum to lattice states.
$\left|[E^+,\pm]_{\text{W}_3}\right\rangle$ is a state that transforms in the $E^+$ irrep of $\text{W}_3$ with a $+1$ eigenvalue
under spatial inversion and $\pm1$ eigenvalue under rotation $R_{12}$.
We identify $\text{SU}(2)_{\text{SW}_3}$ as the diagonal subgroup of
$\text{SU}(2)_S\times\text{SU}(2)_{\mathds{Q}_8}$~\cite{Golterman:1984dn}.
Then, by using the Clebsch-Gordan coefficients, the components are related as
\begin{align}
    &\left| \left[2, 0 \right]_{\text{SW}_3}\right\rangle =\nonumber \\
	&\frac{1}{\sqrt{2}}
	\left(\left| 
	\bigg[ \frac{1}{2}, \frac{1}{2}\bigg]_S 
	\bigg[ \frac{3}{2}, -\frac{1}{2}\bigg]_{\mathds{Q}_8}
	\right\rangle +
	\left| 
	\bigg[ \frac{1}{2}, -\frac{1}{2}\bigg]_S 
	\bigg[ \frac{3}{2}, \frac{1}{2}\bigg]_{\mathds{Q}_8}
	\right\rangle\right),
	\label{eq:diagtocon1}
\end{align}
and
\begin{align}
    &\frac{1}{\sqrt{2}}\bigg(\left| \left[2, 2 \right]_{\text{SW}_3}\right\rangle +
	\left| \left[2, -2 \right]_{\text{SW}_3}\right\rangle\bigg)  = \nonumber \\
	&\frac{1}{\sqrt{2}}
	\left(\left| 
	\bigg[ \frac{1}{2}, \frac{1}{2}\bigg]_S 
	\bigg[ \frac{3}{2}, \frac{3}{2}\bigg]_{\mathds{Q}_8}
	\right\rangle +
	\left| 
	\bigg[ \frac{1}{2}, -\frac{1}{2}\bigg]_S 
	\bigg[ \frac{3}{2}, -\frac{3}{2}\bigg]_{\mathds{Q}_8}
	\right\rangle\right).
	\label{eq:diagtocon2}
\end{align}

Taking all the results of this appendix together, we have
\begin{widetext}
\begin{align}
    \left|16,+\vec{0}\right\rangle \equiv &
    \frac{1}{\sqrt{2}}
	\left(\left| 
	\bigg[ \frac{1}{2}, \frac{1}{2}\bigg]_S 
	\bigg[ \frac{3}{2}, \frac{3}{2}\bigg]_F
	\bigg[ \frac{3}{2}, -\frac{1}{2}\bigg]_{\mathds{Q}_8}
	\bigg[ \frac{1}{2}, \frac{1}{2}\bigg]_{\text{D}_4}
	\right\rangle +
	\left| 
	\bigg[ \frac{1}{2}, -\frac{1}{2}\bigg]_S 
	\bigg[ \frac{3}{2}, \frac{3}{2}\bigg]_F
	\bigg[ \frac{3}{2}, \frac{1}{2}\bigg]_{\mathds{Q}_8}
	\bigg[ \frac{1}{2}, \frac{1}{2}\bigg]_{\text{D}_4}
	\right\rangle\right) \nonumber \\
	&\to 
	\left|\bigg[\frac{3}{2},\frac{3}{2}\bigg]_F\bigg[16, +\vec{0}\bigg]_\text{GTS}\right\rangle ,
	\label{eq:16p}
	\\[1.5em]
	\left|16,-\vec{0}\right\rangle \equiv &
    \frac{1}{\sqrt{2}}
	\left(\left| 
	\bigg[ \frac{1}{2}, \frac{1}{2}\bigg]_S 
	\bigg[ \frac{3}{2}, \frac{3}{2}\bigg]_F
	\bigg[ \frac{3}{2}, \frac{3}{2}\bigg]_{\mathds{Q}_8}
	\bigg[ \frac{1}{2}, \frac{1}{2}\bigg]_{\text{D}_4}
	\right\rangle +
	\left| 
	\bigg[ \frac{1}{2}, -\frac{1}{2}\bigg]_S 
	\bigg[ \frac{3}{2}, \frac{3}{2}\bigg]_F
	\bigg[ \frac{3}{2}, -\frac{3}{2}\bigg]_{\mathds{Q}_8}
	\bigg[ \frac{1}{2}, \frac{1}{2}\bigg]_{\text{D}_4}
	\right\rangle\right)  \nonumber \\
    &\to
    \left|\bigg[\frac{3}{2},\frac{3}{2}\bigg]_F\bigg[16, -\vec{0}\bigg]_\text{GTS}\right\rangle .
	\label{eq:16m}
\end{align}
\end{widetext}
Here, $\left|16,\pm\vec{0}\right\rangle$ have been introduced as shorthand notation for the continuum states for future reference.

\subsection{Quantum numbers of the current operators}

The last ingredient needed for the Wigner-Eckart theorem is the irreducible tensor current operator.
In this work, we use the local isovector axial current, $\Alat$ and local isovector vector current, $\Vlat$, which
have spin-tastes $\gamma_z\gamma_{5}\otimes\xi_z\xi_{5}$ and $\gamma_4\otimes\xi_4$ respectively.
We will need their $\text{SU(2)}_S \times \text{SU}(2)_{F} \times \text{SU}(4)_T$ quantum numbers, just as in the above sections.

The spin and flavor quantum numbers of the currents are straightforward.
By construction, both currents have a total isospin equal to one, with $I_z=0$ components.
$\Alat$ is a spin-1 current with $S_z=0$, and $\Vlat$ is a spin scalar.
The nontrivial part of the identification comes from mapping the quantum numbers of
$\text{SU}(2)_{\mathds{Q}_8}\times\text{SU}(2)_{\text{D}_4}$ to the full $\text{SU}(4)_T$ group.
The quark bilinears we use\footnote{We do not use the taste-scalar current as it is a multilink operator, which has been empirically
observed to have more noise.} transform in the $15$ (adjoint) irrep of $\text{SU}(4)_T$.
The decomposition of the 15 irrep into $\text{SU}(2)_{\mathds{Q}_8}\times\text{SU}(2)_{\text{D}_4}$ irreps occurs via
\begin{equation}
    15 \to (1,1)\oplus(1,0)\oplus(0,1).
    \label{eq:mesonicdecomp}
\end{equation}
Just as above, the quantum numbers of $\text{SU}(2)_{\mathds{Q}_8}\times\text{SU}(2)_{\text{D}_4}$ can label the $15$ irrep of
$\text{SU}(4)_T$ as there are no degenerate irreps in Eq.~(\ref{eq:mesonicdecomp}).
It should be noted that on the lattice, bosonic irreps can be classified according to a subgroup of the GTS group called the
$\overline{\text{RF}}$ group.

We will first examine the continuum quantum numbers of the local lattice vector current, $\Vlat$.
At zero-momentum, it has spin-taste $\gamma_4\otimes\xi_4$.
Within the $\overline{\text{RF}}$ group, $\Vlat$ transforms as the trivial irrep, $1$ \cite{Golterman:1985dz}.
We can decompose $\overline{\text{RF}}$ into the discrete rotational subgroup, $\text{SW}_3$, to get
\begin{equation}
    1 \to A_1,
\end{equation} 
where $A_1$ is the trivial irrep of $\text{SW}_3$. 

We denote as $\Vcont$ the continuum operator corresponding to $\Vlat$ and apply the same subduction procedure as in the previous
session by following the subgroup chain
$\text{SU}(2)_S\times\text{SU}(2)_{\mathds{Q}_8}\to\text{SU}(2)_{\text{SW}_3}\to\text{SW}_3$.
The spin-$0$ irrep of $\text{SU}(2)_{\text{SW}_3}$ subduces into the trivial irrep of $\text{SW}_3$ \cite{Basak:2005ir}.
Consequently, $\Vcont$ needs to be in the trivial irrep of $\text{SU}(2)_S\times \text{SU}(2)_{\mathds{Q}_8}$, and matching
${\mathds{Q}_8}$ factors, $\Vcont$ can only transform as $(0,1)$ irrep of
$\text{SU}(2)_{\mathds{Q}_8}\times\text{SU}(2)_{\text{D}_4}$ from Eq.~(\ref{eq:mesonicdecomp}).

We have just found that $\Vcont$ is a triplet of $\text{SU}(2)_{\text{D}_4}$, and so we need to determine its $z$-component quantum
number.
With positive parity, the three $m^{\text{D}_4}$ components of the $(0,0,1)$ irrep from
$\text{SU}(2)_S\times\text{SU}(2)_{\mathds{Q}_8}\times\text{SU}(2)_{\text{D}_4}$ subduce into the lattice currents
$\gamma_4\otimes\gamma_4$, $\gamma_4\otimes\xi_4\xi_5$ , and $\gamma_4\otimes\xi_5$.
Each transforms trivially in $\overline{\text{RF}}$.
The first lattice current is local and the other two are non-local with multi-link connections between the quarks and antiquarks.
The eigenvalues of $I_S$ are $+1$ for the local current, and $-1$ for the other two.
As discussed in Eq.~(\ref{eq:IS}), the matrix representation of $I_S$ in the continuum can be constructed from the tensor product of
representations of $\text{SU}(2)_{\text{D}_4}$, $U(1)_{\text{D}_4}$, and $P$ to give
\begin{align}
    e^{i\pi\times\text{diag}(1,0,-1)}\otimes 1\otimes 1 = 
        \begin{bmatrix}
            -1 & 0 & 0 \\
            0 & 1 & 0 \\
            0 & 0 & -1 \\
        \end{bmatrix} = I_S,
    \label{eq:ISMatrix_gV}
\end{align}
where the $\text{SU}(2)_{\text{D}_4}$ factor is in a spin triplet as discussed, $U(1)_{\text{D}_4}$ is a trivial factor to give the
correct $I_S$ eigenvalues, and the parity is also trivial by construction.
Consequently, to get the correct $I_S=1$ eigenvalue on the lattice, the local $\gamma_4\otimes\gamma_4$ current must have zero
$z$-component in the triplet irrep of $\text{SU}(2)_{\text{D}_4}$ in the continuum limit.
This completes the subduction of $\Vcont$ into $\Vlat$.

The procedure is similar subducing the continuum axial-vector current $\Acont$ into the lattice version $\Alat$.
On the lattice, $\Alat$ transforms as a three-dimensional irrep, $3''''$, of $\overline{\text{RF}}$, which decomposes into
the
\begin{align}
    3'''' \to A_1 \oplus E
\end{align}
irreps of $\text{SW}_3$. The linear combination
\begin{align}
    A_1 \propto  (\gamma_x\gamma_5\otimes\xi_x\xi_5)+(\gamma_y\gamma_5\otimes\xi_y\xi_5)+(\gamma_z\gamma_5\otimes\xi_z\xi_5)
\end{align}
transforms trivially under discrete rotations so it lives in the $A_1$ irrep.
The remaining linear combinations are
\begin{align}
	E_+&\propto(\gamma_x\gamma_5\otimes\xi_x\xi_5) +( \gamma_y\gamma_5\otimes\xi_y\xi_5) -2(\gamma_z\gamma_5\otimes\xi_z\xi_5) \\
    E_-&\propto(\gamma_x\gamma_5\otimes\xi_x\xi_5) -( \gamma_y\gamma_5\otimes\xi_y\xi_5),
\end{align}
where the subscript on the left-hand side is the eigenvalue $\pm$ of~$R_{12}$.

In the continuum, $\Acont$ is a spin-$1$ operator of $\text{SU}(2)_S$.
The $A_1$ irrep subduces from the spin-$0$ irrep of $\text{SU}(2)_{\text{SW}_3}$ and the $E$ irrep subduces from the spin-$2$ irrep
of $\text{SU}(2)_{\text{SW}_3}$.
With the rules for the addition of angular momentum, this requires $\Acont$ to be in the irrep (1,1) of $\text{SU}(2)_S\times
\text{SU}(2)_{\mathds{Q}_8}$ with zero $z$-component spins in both $\text{SU}(2)$ factors.

Now, according to Eq.~(\ref{eq:mesonicdecomp}), $\Acont$ can either be a spin-$0$ or $1$ operator of $\text{SU}(2)_{\text{D}_4}$.
Recall that on the lattice, $\text{D}_4$ is generated by the transformations $I_S$ and $\Xi_{123}$ \cite{Lin:2019pia}.
$\Alat$ is an eigenvector of both these symmetries with respective eigenvalues $1$ and $-1$.
Because $\text{SU}(2)_{\text{D}_4}$ subduces into the $\text{D}_4$ factor of the GTS group, these non-trivial eigenvalues mean that
$\Acont$ cannot transform trivially under $\text{SU}(2)_{\text{D}_4}$.
As such, $\Acont$ can only belong to spin-$1$ irrep of $\text{SU}(2)_{\text{D}_4}$.
Further, it has zero $z$-component following the same argument in Eq.~(\ref{eq:ISMatrix_gV}).

In summary, we have determined the continuum quantum numbers of $\Acont$ and $\Vcont$, which subduce into the desired lattice
current operators, $\Alat$ and $\Vlat$ respectively.
Using the same notation as in Eq.~(\ref{eq:conket}), the continuum currents transform as

\begin{align}
    -\Acont^{(1,0)_S(1,0)_F}_{(1,0)_\mathds{Q}(1,0)_{\text{D}_4}}& \equiv \Acont/\sqrt{n_t} \to \Alat/\sqrt{n_t},
    \label{eq:jA} \\
    \Vcont^{(0,0)_S(1,0)_F}_{(0,0)_\mathds{Q}(1,0)_{\text{D}_4}} & \equiv \Vcont/\sqrt{n_t} \to \Vlat/\sqrt{n_t}.
\label{eq:jV}
\end{align}

The spin and flavor quantum numbers of the tensor operators are denoted by the superscripts, whereas the taste quantum numbers are
given in the subscripts.
$n_t=4$ is the number of tastes and $\sqrt{n_t} = 2$ is required to properly normalize tensor operators.
The minus sign in front of the axial current is a convention that we follow according to Table I of Ref.~\cite{Hecht:1969ck}.

As an aside, there is an easy way to obtain the continuum taste quantum numbers of an arbitrary quark bilinear without explicit
group subduction.
Table I of Ref.~\cite{Hecht:1969ck} outlines the $\text{SU}(4)$ generators and their corresponding tensor operators.
Once we adopt the Euclidean Dirac representation for the taste gamma matrices $\xi_4 = \sigma_3\otimes I, \xi_j = \sigma_2\otimes
\sigma_j$ (where $\sigma_j$ are the usual Pauli matrices), those generators give the components of the continuum taste matrices.
For example, the local axial and vector currents we use have taste gamma matrices of
\begin{align}
    \xi_z\xi_5 &= 
        \left[\begin{array}{cccc}
        1&0&0&0\\
        0&-1&0&0\\
        0&0&-1&0\\
        0&0&0&1\\
        \end{array}
        \right]\\
        \xi_4 &= 
        \left[\begin{array}{cccc}
        1&0&0&0\\
        0&1&0&0\\
        0&0&-1&0\\
        0&0&0&-1\\
        \end{array}
        \right].
\end{align}
They are proportional to the generators $\frac{1}{2}(A_{11}-A_{22}-A_{33}+A_{44})$ and $\frac{1}{2}(A_{11}+A_{22}-A_{33}-A_{44})$.
By identifying $S$ as $\text{D}_4$ in Table I of Ref.~\cite{Hecht:1969ck}, and similarly $T$ as $\mathds{Q}_8$, we can recognize the
tensor product $S\otimes T = \sigma_3\otimes\sigma_3$ and $\sigma_3\otimes I$, indicating a spin-1 representation whenever a
$\sigma_3$ appears in the tensor product.
This yields the continuum taste quantum numbers of these states as $(1,0)_{\mathds{Q_8}}(1,0)_{\text{D}_4}$ and
$(0,0)_{\mathds{Q}_8}(1,0)_{\text{D}_4}$.%
\footnote{There is a typo in Table~I of Ref.~\cite{Hecht:1969ck}.
The irreducible tensor components at line~3 should read $-T^{[211]}_{(1,0)(1,0)}$ instead of $-T^{[211]}_{(0,0)(0,0)}$.}
 
\section{Wigner-Eckart Theorem and the Physical Matrix Elements}
\label{app:WET}

In this appendix we need to relate, for each current, the $s=\pm 0$ nucleon-like lattice matrix elements to their physical continuum
counterpart.
We label the continuum matrix elements as
\begin{align}
&M^V_\pm \equiv \left\langle16,\pm\vec{0}| \Vcont|16,\pm\vec{0}\right\rangle\label{eq:unphygV},\\
&M^A_\pm \equiv \left\langle16,\pm\vec{0}| \Acont|16,\pm\vec{0}\right\rangle\label{eq:unphygA}.
\end{align}

Since we know the continuum quantum numbers of each state and current, we can apply the Wigner-Eckart theorem to relate the
different components.
To further reduce the number of independent matrix elements from four to two, we apply the Wigner-Eckart theorem to the
$\text{SU}(2)_{\mathds{Q}_8}$ part of the irreps in Eqs.~(\ref{eq:16p}), (\ref{eq:16m}), (\ref{eq:jA}), and (\ref{eq:jV}) to find
\begin{align}
&M_-^A = -3M_+^A, \label{eq:mA}\\
&M_-^V = M_+^V. 
\end{align}

This result is consistent with the discussion around Appendix~(\ref{app:nucleon_on_lattice}).
On the lattice, we have found exact symmetries for the local vector currents
\begin{align}
    \bigg\langle\bigg[\frac{3}{2},\frac{3}{2}\bigg]_F\bigg[16,+\vec{0}\bigg]_\text{GTS}
    \bigg|
    \Vlat
    \bigg|
    \bigg[\frac{3}{2},\frac{3}{2}\bigg]_F\bigg[16, +\vec{0}\bigg]_\text{GTS}\bigg\rangle = \nonumber\\
    \bigg\langle\bigg[\frac{3}{2},\frac{3}{2}\bigg]_F\bigg[16, -\vec{0}\bigg]_\text{GTS}
    \bigg|
    \Vlat
    \bigg|
    \bigg[\frac{3}{2},\frac{3}{2}\bigg]_F\bigg[16, -\vec{0}\bigg]_\text{GTS}\bigg\rangle ,
    \label{eq:gvlatt}
\end{align}
which comes from Eq.~(\ref{eq:2ptrelation}) and
\begin{align}
    &\sum_{\vec{x},\vec{y}} \left\langle B^{16}_{+\vec{0}}(\vec{x},t)\Vlat(\vec{y},\tau)
        \overline{B}^{16}_{+\vec{0}}(0)\right\rangle = \nonumber\\
    &~~~~~~~~~~~~~~~~~~\sum_{\vec{x},\vec{y}} \left\langle B^{16}_{-\vec{0}}(\vec{x},t)\Vlat(\vec{y},\tau)
        \overline{B}^{16}_{-\vec{0}}(0)\right\rangle , 
\label{eq:gvcorr_exact}
\end{align}
derived from applying lattice rotations and shifts.
For the local axial-vector current, there are no symmetries relating the matrix elements on the lattice, but the relationship in
Eq.~(\ref{eq:mA}) emerges in the continuum.

\begin{figure}
	\centering
	\includegraphics[width=0.98\columnwidth]{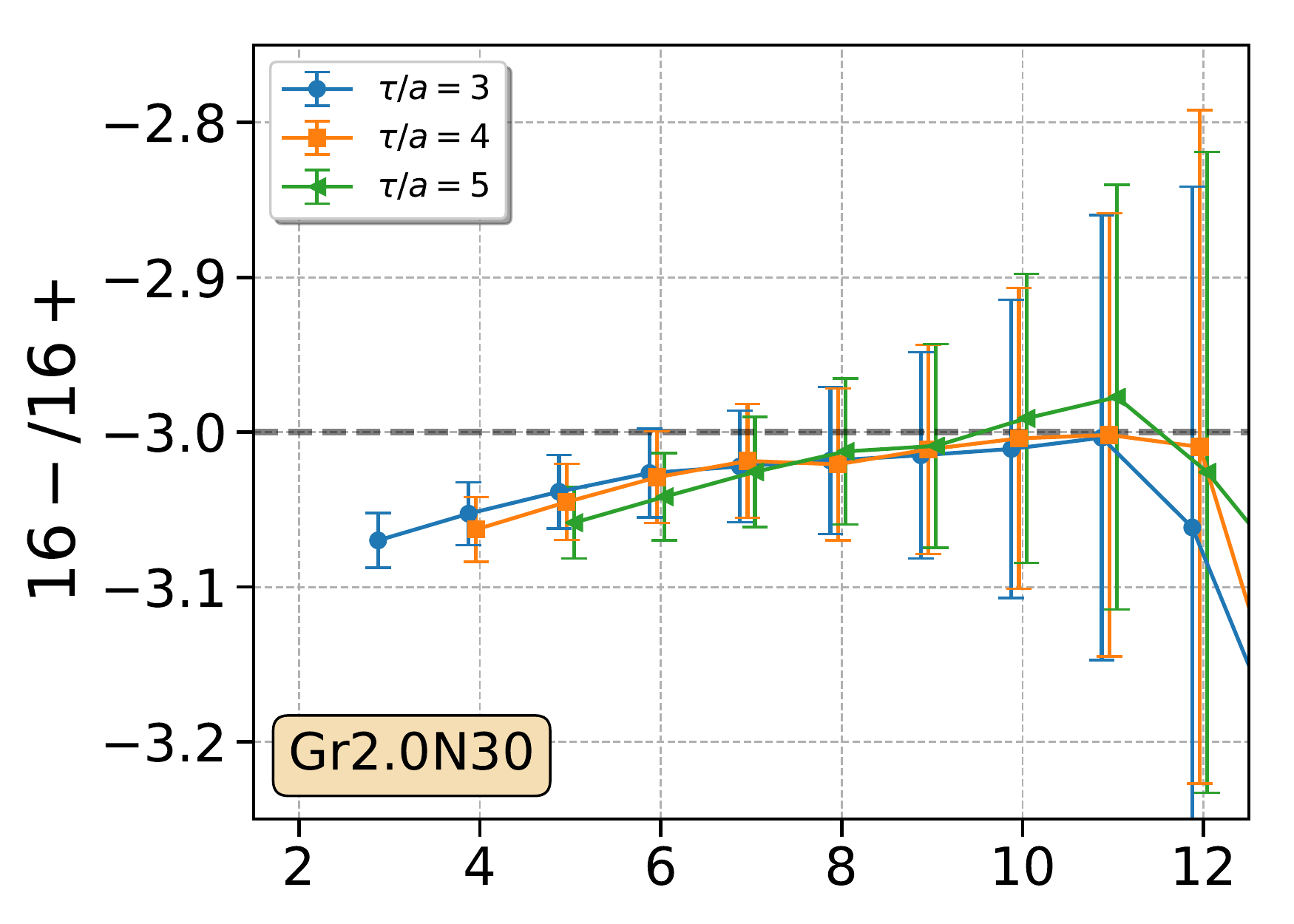}
	\includegraphics[width=0.98\columnwidth]{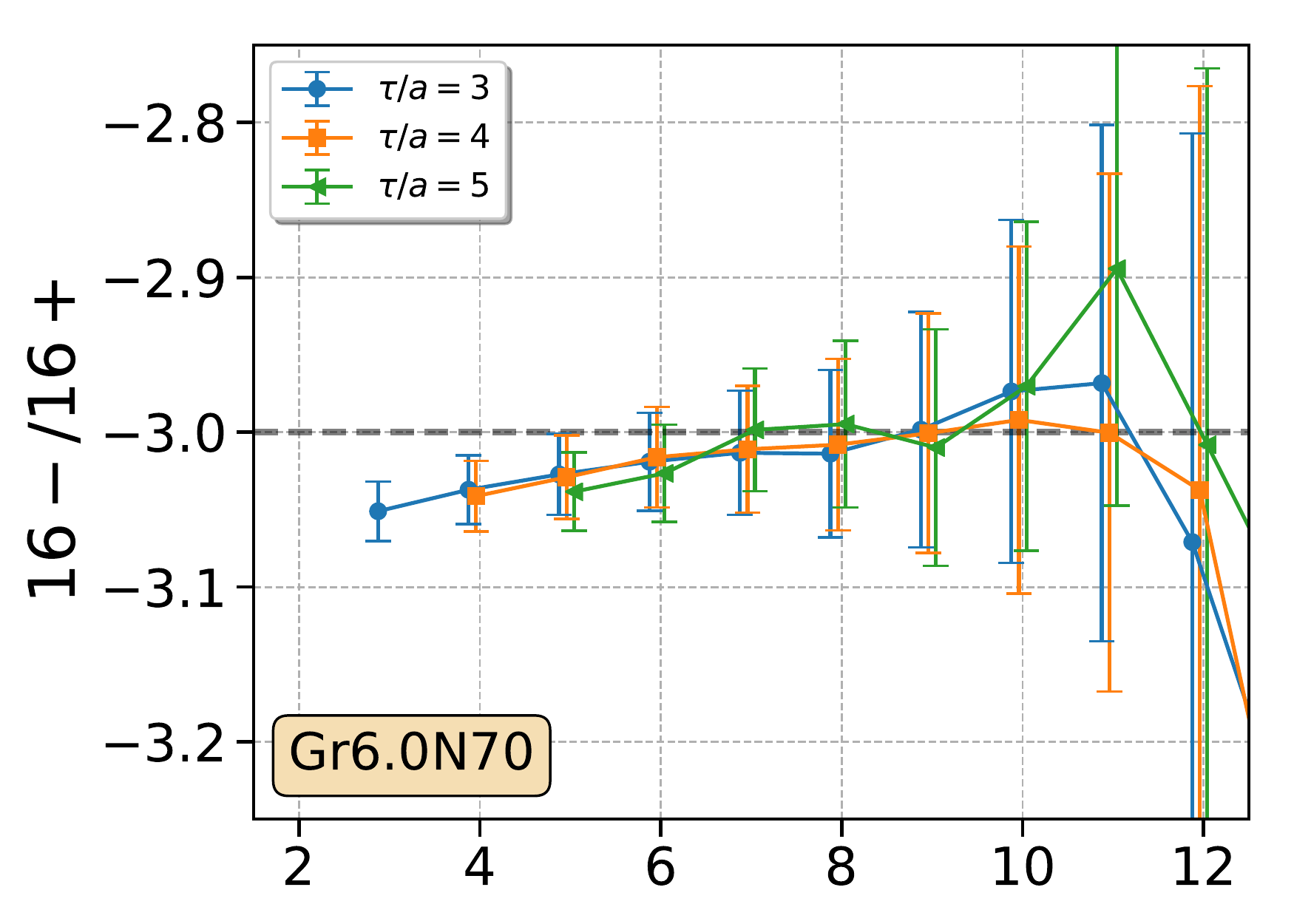}
	\caption{(Color online) The ratio of the three-point $g_A$ correlators, built with $\sum_{\vec{x}}B^{16}_{-\vec{0}}(\vec{x},t)$
        ("16-" with eigenvalues of $-1$ for the $x-y$ plane rotation) and  $\sum_{\vec{x}}B^{16}_{+\vec{0}}(\vec{x},t)$ ("16+" with 
        eigenvalue of $+1$) interpolating operators, as a function of source-sink separation time $t$ and current insertion time 
        $\tau$.
        Both interpolators are used in Eq.~(\ref{eq:3ptraw_gA}) to compute $g_A$. Each plot represents a different Wuppertal
        smearing at the sink, with parameters  $0.2$ (Gr2.0N30) and $0.6$fm (Gr6.0N70) rms radii.
        The group theory requires that in the continuum and $t,\tau\to\infty$ limits that the ratio is equal to $-3$,
        which is shown as a dashed lines.}
	\label{fig:16m16p_ratio}
\end{figure}

To demonstrate this observation, we have plotted the ratio of optimized $g_A$ three-point correlators created with
$\sum_{\vec{x}}B^{16}_{-\vec{0}}(\vec{x},t)$ and $\sum_{\vec{x}}B^{16}_{+\vec{0}}(\vec{x},t)$ interpolators in
Fig.~\ref{fig:16m16p_ratio}.
In the limits $\tau,t-\tau\to\infty$ and $a\to 0$, the ratio should converge to the dashed lines at $-3$ as predicted
by the above group theory.
The small deviation is caused by a combination of excited state contamination, discretization effects, and taste-breaking effects.
The same ratio for the vector current is consistent with one to high precision, as enforced by the lattice relation in
Eq.~(\ref{eq:gvcorr_exact}).
Fig.~\ref{fig:16m16p_ratio} is therefore a non-trivial verification of our group theory understanding of staggered baryon matrix
elements.

To relate the two remaining staggered matrix elements to their counterparts in QCD without tastes we observe that 
\begin{align}
    |M_{\text{phy}}^V| &= |\left\langle B|\Vcont|B\right\rangle|\label{eq:phygV},\\
    |M_{\text{phy}}^A| &= |\left\langle B|\Acont|B\right\rangle|\label{eq:phygA},
\end{align}
where $M_{\text{phy}}^V$ and $M_{\text{phy}}^A$ are the physical vector and axial matrix elements. Here 
\begin{align}
	\left| B\right\rangle \equiv  
	\left| 
	\bigg[ \frac{1}{2}, \frac{1}{2}\bigg]_S 
	\bigg[ \frac{1}{2}, \frac{1}{2}\bigg]_F
	\bigg[ \frac{3}{2}, \frac{3}{2}\bigg]_{\mathds{Q}_8}
	\bigg[ \frac{3}{2}, \frac{3}{2}\bigg]_{\text{D}_4}
	\right\rangle
	\label{eq:single_taste}
\end{align}
is the single-taste nucleon, e.g., $|B\rangle$ has the correct isospin of $\frac{1}{2}$ and transforms as the symmetric $20_S$ irrep
of $\text{SU}(4)_T$.
The $20_S$ irrep of $\text{SU}(4)_T$ contains states with a single-taste baryon\footnote{ As an analogy, the single-taste of
$\text{SU}(2)_T$ is similar to the $\Delta^{++}$ (consisting of three valence up-quarks) in $\text{SU}(2)_F$}.
To relate single-taste baryon matrix elements to the physical one, we also need the taste-diagonal current operators which have
tastes $\xi_z\xi_5$, $\xi_4$, $\xi_1\xi_2$, or $\mathbf{1}$.
These constructions must coincide with the physical matrix elements, up to a sign, if the taste restoration is valid in the
continuum limit.

Again, we can use the quantum numbers of $\text{SU}(2)_{\mathds{Q}_8}\times \text{SU}(2)_{\text{D}_4}$ to uniquely label components
in $20_S$ because there are no degenerate irreps in the decomposition $20_S\to \left(\frac{3}{2},
\frac{3}{2}\right)\oplus\left(\frac{1}{2}, \frac{1}{2}\right)$.
We apply the Wigner-Eckart theorem to normalize the matrix elements, $M_-^V$ and $M_-^A$, to $M_{\text{phy}}^V$ and
$M_{\text{phy}}^A$.
This boils down to finding the correct Clebsch-Gordon coefficients to rotate $|16,-\vec{0}\rangle$ to the single-taste baryon
$|B\rangle$ while leaving the taste-diagonal currents unchanged.
An $\text{SU}(4)_T$ rotation alone is insufficient because these states belong to different $\text{SU}(4)_T$ irreps.
However we can embed flavor and taste into a larger group and perform rotations in this larger group to accomplish the task.
If we consider the relevant group factors $\text{SU}(4)_{F\times \text{D}_4}\supset\text{SU}(2)_{F}\times\text{SU}(2)_{\text{D}_4}$,
both $|16,-\vec{0}\rangle$ and $|B\rangle$ belong to the same $20_M$ irrep of $\text{SU}(4)_{F\times \text{D}_4}$, and so we can
apply the Wigner-Eckart theorem to this group.

The details of the generalized Wigner-Eckart theorem for $\text{SU}(4)$ are described in Ref.~\cite{Hecht:1969ck}.
We will only need the Wigner-Eckart theorem in Eq.~(33) of that reference, and the Clebsch-Gordon coefficients in Table A4.5 of
Ref.~\cite{Hecht:1969ck}, to conclude that
\begin{align}
    |M_-^V| &= |\langle B|\Vcont|B\rangle| = |M_{\text{phy}}^V| \\
    |M_-^A| &= |\langle B|\Acont|B\rangle| = |M_{\text{phy}}^A|.
    \label{eqn:mphys}
\end{align}

We can understand the trivial normalization factor by realizing that in the continuum, $\text{SU}(2)_F$,
$\text{SU}(2)_{\mathds{Q}_8}$, and $\text{SU}(2)_{\text{D}_4}$ are indistinguishable from one another because of the enlarged
$\text{SU}(8)_{FT}$ symmetry.
This means that the matrix elements are invariant under the exchange of $\text{D}_4$ and $F$ labels in Eq.~(\ref{eq:phygV}) and
Eq.~(\ref{eq:phygA}).
This shows that Eq.~(\ref{eq:phygV}) and Eq.~(\ref{eq:phygA}) are identical to Eq.~(\ref{eq:unphygV}) and Eq.~(\ref{eq:unphygA}),
and hence, the trivial normalization factors.
Combining the shift symmetry relationship in the correlators from Eqs.~(\ref{eq:threepointcorrelator}) with Eqs.~(\ref{eqn:mphys})
gives a key result for this paper, which is presented in Eq.~(\ref{eq:3ptraw_gV}), (\ref{eq:3ptraw_gA}).

\acknowledgments

We are grateful to the MILC collaboration for the use of the source code adapted to generate the correlators in this study and for
permission to use their 2+1+1-flavor gauge-field ensemble.
Computations for this work were carried out on facilities of the USQCD Collaboration, which are funded by the Office of Science of
the U.~S.\ Department of Energy.
This manuscript has been authored by Fermi Research Alliance, LLC under Contract No.~DE-AC02-07CH11359 with the U.~S.\ Department of
Energy, Office of Science, Office of High Energy Physics.
Brookhaven National Laboratory is supported by the U.~S.\ Department of Energy under Contract No.~DE-SC0012704.
Additional support was provided under U.~S.\ DOE Contract No. DE-SC00190193. 

\bibliographystyle{apsrev4-1}
\bibliography{main,usqcd-wp}

\end{document}